\documentclass[a4paper,12pt]{article}
\usepackage[utf8]{inputenc}
\usepackage{geometry}
\usepackage[sumlimits,intlimits,namelimits]{amsmath}
\usepackage{empheq}
\usepackage{amssymb}
\usepackage{a4wide}
\usepackage[english]{babel}
\usepackage{hyperref}
\usepackage{longtable, multirow,tabularx}
\usepackage[titletoc,title]{appendix}
\usepackage{slashed}
\usepackage{amsmath}
\usepackage{cite}
\DeclareMathOperator{\Tr}{Tr}
\makeatletter
\def\fmslash{\@ifnextchar[{\fmsl@sh}{\fmsl@sh[0mu]}}
\def\fmsl@sh[#1]#2{%
  \mathchoice
    {\@fmsl@sh\displaystyle{#1}{#2}}%
    {\@fmsl@sh\textstyle{#1}{#2}}%
    {\@fmsl@sh\scriptstyle{#1}{#2}}%
    {\@fmsl@sh\scriptscriptstyle{#1}{#2}}}
\def\@fmsl@sh#1#2#3{\m@th\ooalign{$\hfil#1\mkern#2/\hfil$\crcr$#1#3$}}
\makeatother
\usepackage{graphicx}
\usepackage{subfigure}
\usepackage{epstopdf}
\usepackage{xcolor}

\def\Li{\mbox{Li}}
\def\L{\mbox{L}_{\mbox{\scriptsize R}}}

\newcommand{\ice}[1]{\relax}

\begin{document}
\begin{titlepage}
\begin{flushright}
SI-HEP-2021-11 \\[0.2cm]
SFB-257-P3H-21-020
\end{flushright}

\vspace{1.2cm}
\begin{center}
{\Large\bf
Master Integrals for Inclusive Weak Decays \\[2mm]
of Heavy Flavours at Next-to-Leading Order  

%The differential $B$-meson semi-leptonic width at NLO\\ 
}
\end{center}

\vspace{0.5cm}
\begin{center}
{\sc Th.~Mannel, D.~Moreno and A.~A.~Pivovarov} \\[2mm]
{Center for Particle Physics Siegen, Theoretische Physik 1, Universit\"at Siegen\\ 57068 Siegen, Germany}  
\end{center}

\vspace{0.8cm}
\begin{abstract}
\vspace{0.2cm}\noindent
We present analytical results for master integrals emerging in 
the computation of differential rates for 
inclusive weak decays of heavy flavors 
at next-to-leading order 
(NLO) in QCD.
As an immediate physical application,
these master integrals allow for a calculation of the spectra of the 
leptonic invariant mass in inclusive semileptonic decays 
in the framework of the heavy quark 
expansion, including the NLO QCD corrections to power suppressed terms.

%For  semi-leptonic decays this allows for the calculation
%of the width differential in the lepton pair invariant mass
%analytically.
%The new setup allows for a systematic computation of power suppressed
%terms
%within HQE with $\alpha_s$ corrections.
%Our method amounts to computation of the imaginary 
%part of two-loop integrals with three different masses.
 
%present a new approach based on the use of the optical theorem and the heavy quark expansion to compute the inclusive $B$-hadron semi-leptonic decay 
%width differential in the lepton pair invariant mass analytically. The key feature is the use of the dispersion representation of the leptonic loop.
%The new setup allows for a systematic computation of power and $\alpha_s$ corrections and the possibility to compute moments of the distribution 
%with cuts. We develop our method for the computation of $\alpha_s$ corrections at NLO, which requires the computation of the imaginary 
%part of two-loop integrals with two different scales.

%moments are used for the precise detrmination of the HQE parameters. 
\end{abstract}

\end{titlepage}

\newpage
\pagenumbering{arabic}
%

%
%\tableofcontents
%
\newpage

%\section{Structure}

%1. How important differential rate is ... (bla-....)
%2. Formulation
%3. Masters
%4. Results
%5. Discussion of results in massless limit as the formulas in this case
%are much shorter and viewable.
%6. Conclusion
%7. Literature

\section{Introduction}

The Heavy Quark Expansion (HQE)~\cite{Chay:1990da,Bigi:1992su,Manohar:1993qn,Mannel:1993su} 
provides a solid framework for the computation of observables for hadrons containing a heavy quark. 
For this reason it has become the standard tool to analyze inclusive heavy hadron decays, which 
has been refined over the last three decades by detailed calculations of higher order 
perturbative contributions, as well as by inclusion of nonperturbative contributions. 
The techniques for the computation of inclusive semi-leptonic widths and moments of the 
distribution have been developed long ago~\cite{Falk:1995me,Bauer:2002va,Voloshin:1994cy}. 
However, the fully differential width at order $\alpha_s$ was obtained only fifteen years ago~\cite{Trott:2004xc,Uraltsev:2004in,Aquila:2005hq}, including 
double and triple differential rates and spectral moments, some of which could be computed 
even analytically.

From the phenomenological side, inclusive decays are of great interest for the precise 
extraction of Cabibbo-Kobayashi-Maskawa (CKM) matrix elements~\cite{Gambino:2007rp,Capdevila:2021vkf},
which play a central role for testing the flavor sector of the Standard Model of Particle Physics (SM) 
and for the search for flavour and CP violation beyond the SM. 

In this paper we describe a new setup for the computation of $\alpha_s$ corrections to 
differential rates in some specific kinematic invariants. In particular, this setup is 
designed such that it can be easily extended to the calculation of NLO QCD corrections to 
power-suppressed terms within the HQE. To this end, it opens the road for analytic NLO calculations 
for differential rates for power-supressed terms, which is a pheonemenologically important, since 
this allows us to implement phase-space cuts in an easy way. 

%Most importantly, the new setup is such that it contains all necessary ingredients 
%for the computation of $\alpha_s$ corrections to power 
%corrections. The differential rate is of great importance as it allows to compute moments with phase space cuts in the corresponding kinematic 
%invariant. That is crucial when comparing to experiment, where phase space cuts are unavoidable. 

The approach makes use of modern tools for the computation of 
multi-loop Feynman integrals, in particular the program LiteRed~\cite{Lee:2012cn,Lee:2013mka}.
%based on the integration-by-parts (IBP) reduction technique. 
That allows to write the Feynman diagram amplitudes as a combination of 
master integrals, which are our key ingredients. These are up to two loop integrals 
with two scales which we compute analytically, meaning in terms of well-known and well-studied special functions. 
Even though inclusive semi-leptonic decays are the main target for phenomenological 
applications, the master integrals 
we compute here can be applied to other inclusive decays, like non-leptonic or exotic decays.

In particular, our formulation allows us to calculate the total rate with final states with three
different masses even for the power suppressed terms.
At leading power the result is known~\cite{Hokim:1983yt} and we use this result as a check 
of our calculation. As an application, we compute for the first time the semi-leptonic differential 
width in the lepton pair invariant mass square with massive charm and 
massless leptons analytically. 

We use standard renormalization and dimensional regularization ($D= 4-2\epsilon$)~\cite{tHooft:1972tcz} 
with anticommuting $\gamma_5$~\cite{Chetyrkin:1997gb,Grozin:2017uto}. Therefore, 
the Dirac algebra of $\gamma$-matrices usually defined in $D=4$ needs to be extended to 
$D$-dimensional spacetime~\cite{Altarelli:1980fi,Buras:1989xd,Buras:1990fn,Chanowitz:1979zu}. 
The Dirac and Lorentz algebra is manipulated using Tracer~\cite{Jamin:1991dp} 
and the $\epsilon$ expansion of Hypergeometric functions is computed with the help of HypExp~\cite{Huber:2005yg,Huber:2007dx}.

The paper is organized as follows. In Sec.~\ref{Sec:formulation} we develop our approach for the computation of the differential decay 
width. In Sec.~\ref{masterint} we compute the necessary master integrals. In Sec.~\ref{Sec:applications} we apply the results of 
Sec.~\ref{masterint} to the computation of the semi-leptonic width differential in the lepton pair invariant mass square, as 
well as the total semi-leptonic width.
Finally, we present the results for the HQE coefficients of the differential and total semi-leptonic decay 
width in Sec.~\ref{Sec:results}. We also give some technical results in the Appendix.

\section{Formulation of the method}
\label{Sec:formulation}

We consider the inclusive decay of a heavy flavoured hadron, which is induced by the 
quark-level transition
\begin{equation}
 Q(M)\rightarrow q(m)p_1(m_1)\ldots p_n(m_n)\,,
\end{equation}
where $Q$ is a heavy quark of mass $M$, %with on-shell external momentum $p^2=M^2$, 
$q$ is a lighter quark of mass $m$ and $p_i$ stands for any type of particles 
with mass $m_i$. In typical Standard Model (SM) processes the $p_i$ are either quarks or leptons, but in fact they can be of any type (fermions, scalars...), i.e. our results are very general.  
%appearing in a hypothetical beyond the Standard Model (BSM) interaction Hamiltonian. 
%With these general definitions we want to stress the generality of the results we obtain.

Using the optical theorem the decay width of a heavy-flavoured hadron can be computed from the discontinuity of the forward scattering matrix element, for which the HQE can be employed, 
yielding a systematic expansion  in powers of $\Lambda_{\mbox{\scriptsize QCD}}/M\ll 1$: 
\begin{eqnarray}
 \Gamma(Q\rightarrow q p_1\ldots p_n) &=& 
 \frac{1}{M_{H_Q}} \langle H_Q(p_{H_Q})\lvert \mbox{Im}\, \hat{T} \lvert H_Q(p_{H_Q})\rangle \nonumber
 \\
 &=& \Gamma^0 \Big( C_0(\rho,\eta_i) +\mathcal{O}(1/M)\Big)
 \,,
\end{eqnarray}
where $\lvert H_Q(p_{H_Q}) \rangle$ stands for the full hadronic states 
with four-momentum $p_{H_Q}$ and mass $M_{H_Q}$, the transition operator $\hat{T}$ is related to the scattering 
operator $\hat{S}$ by $\hat{S} = 1 + i\hat{T}$, the normalization factor $\Gamma^0$ contains all trivial information like 
phase space factors and dimensionful scales and couplings, and $C_0$ is the leading order matching coefficient in the HQE, 
which depends on the dimensionless parameters $\rho=m^2/M^2$ and $\eta_i = m_i^2/M^2$ ($i=1,\ldots,n$).
Note that we assume that $\rho$ and $\eta_i$ are treated as parameters of order unity.  

\begin{figure}[!htb]  
	\centering
	\includegraphics[width=0.7\textwidth]{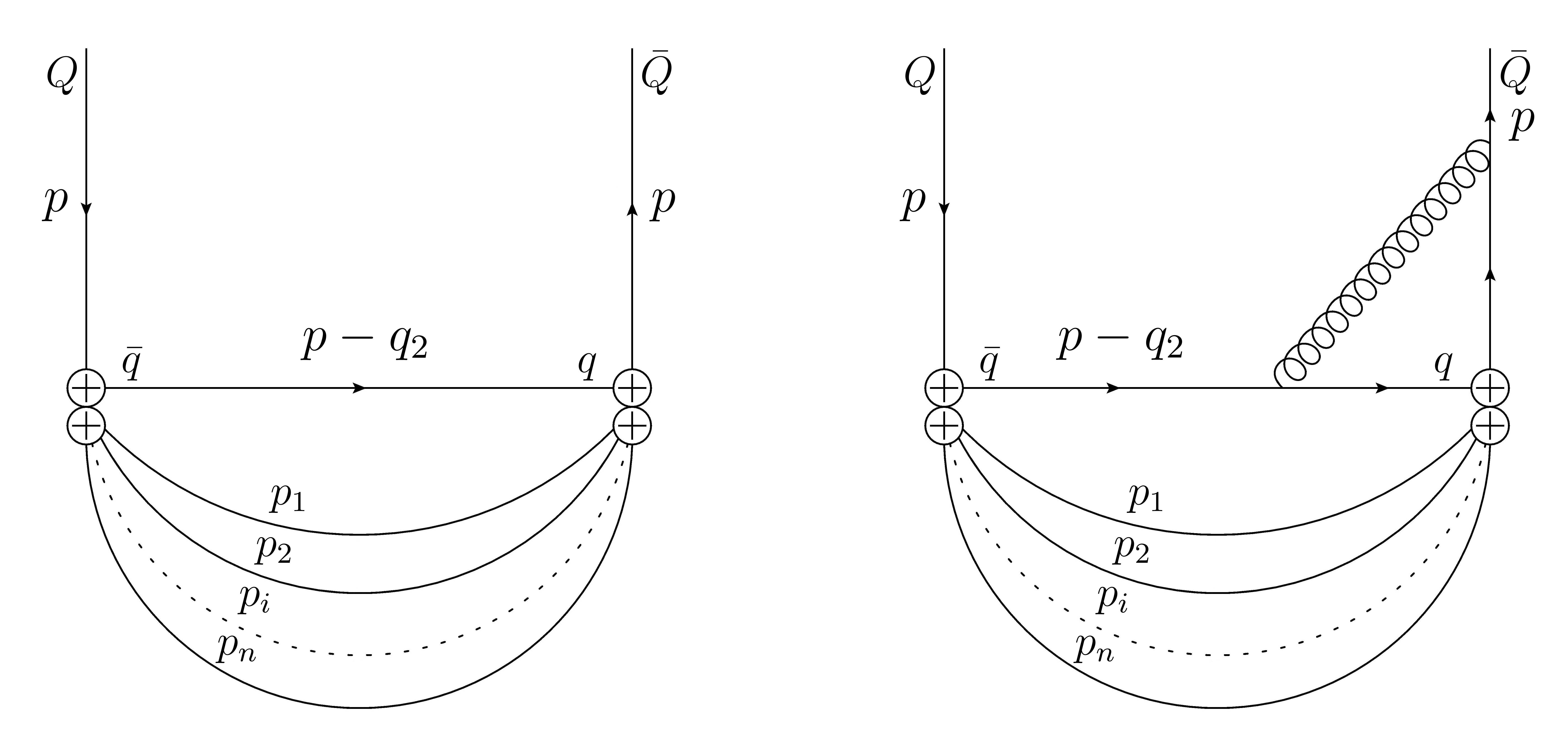}
        \caption{Examples of Feynman diagrams entering in the calculation of the coefficient of the total width $C_0$ at 
        LO-QCD (left) and NLO-QCD (right).}
        \label{GtotGEN}
\end{figure}

The typical Feynman diagram one faces in the computation of the total width at LO-QCD is the $n$-loop graph shown in Fig.~[\ref{GtotGEN}] (left). 
%with on-shell external momentum $p^2=M^2$. 
At NLO-QCD, the required Feynman diagrams are $(n+1)$-loop graphs which can be obtained from the one in Fig.~[\ref{GtotGEN}] (left) by 
taking one gluon exchange (massless line exchanges) between two massive lines. An example of such diagrams is shown in Fig.~[\ref{GtotGEN}] (right). 
To keep the following discussion simple, let us consider colorless particles $p_i$, such that the gluon 
couples only to the quarks $Q$ and $q$. At NLO, this is no restriction of generality.
 
However, our goal is not the computation of the total width, but the width differential 
in some kinematic invariant, which we chose to be  
the invariant mass square of the $n$ $p$-particles. 
Technically, one can achieve this by factorizing the $n$ massive 
$p$-particle propagators in a single massive propagator using a dispersion representation

\begin{figure}[!htb]  
	\centering
	\includegraphics[width=0.8\textwidth]{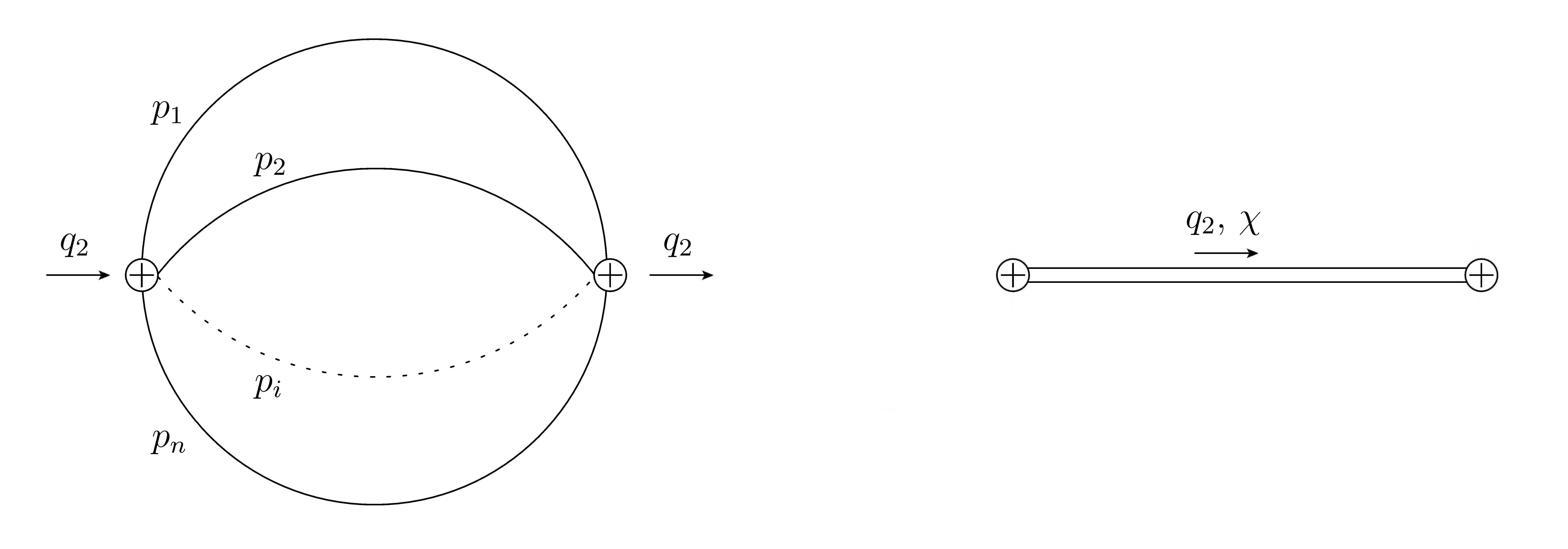}
        \caption{Factorization of the $n$-loop sunset with all massive lines (left) in a 
        single ``effective massive propagator'' with spectral density, mass $\chi$ and
        integration over the $n$ $p$-particle invariant mass square (right).}
        \label{Factorized}
\end{figure}

\begin{equation}
 \label{disp1LSm1m2Generic}
 \frac{1}{i}\int
 \prod_{j=1}^{n} \bigg( \frac{d^D k_j}{(2\pi)^D} \frac{1}{k_j^2-m_j^2} \bigg) \delta^{(4)}\bigg(q_2 - \sum_{i=1}^{n} k_n\bigg)
 = \int_{(\sum_{i=1}^n m_i)^2}^{\infty}  \frac{\rho_{{\scriptsize\mbox{($n-1$)LS}}}(\chi^2,m_1,\ldots,m_n)}{\chi^2 - q_2^2-i\eta}d(\chi^2)\,,
\end{equation}
where $\rho_{{\scriptsize\mbox{($n-1$)LS}}}$ is the spectral density for the ($n-1$)-loop generalized sunset diagram. For example, 
at one-loop it is given by Eq.~(\ref{specden1LSm1m2}). Note that
the $(n-1)$-loop integral in Eq.~(\ref{disp1LSm1m2Generic}) has become an integral over the mass of an ``effective massive propagator'' of mass $\chi$, 
and that the dependence on the $m_i$ is completely factorized in the spectral density. 
This fact is diagramatically represented in Fig.~\ref{Factorized}.
This representation serves as a powerful tool for both, the computation of the total width, and 
the computation of the decay width differential in the $n$ $p$-particle invariant mass square $q_2^2 = (\sum_{i=1}^n k_i)^2$.
%In those cases where the appearance of too many scales makes the analytical integration impossible, this representation allows to 
%write the integrals in a form ready for numerical integration.
The $(n-1)$-loop in Eq.~(\ref{disp1LSm1m2Generic}) is always cut, so $q_2^2$ is forced to be on-shell $q_2^2 = \chi^2$, corresponding to 
the $q_2^2$ pole on left hand side of Eq.~(\ref{disp1LSm1m2Generic}). Therefore, the integrand is differential in the $n$ $p$-particle invariant mass 
square $\chi^2$. 
 
In this setup, the decay width can be written as
\begin{eqnarray}
  \Gamma(Q\rightarrow q p_1\ldots p_n) &=&  \nonumber
  \Gamma^0 \int_{(\sum_{i=1}^n \sqrt{\eta_i})^2}^{(1-\sqrt{\rho})^2}dr\,\rho_s(r,\eta_i)\Big( \mathcal{C}_0(r,\rho) + \mathcal{O}(1/M)\Big)
 \\
 &\equiv& \int_{(\sum_{i=1}^n \sqrt{\eta_i})^2}^{(1-\sqrt{\rho})^2}dr\, \frac{d\Gamma(r,\rho,\eta_i)}{dr} \,,
\end{eqnarray}
where $r = \chi^2/M^2$ is the dimensionless invariant mass square of the $n$ $p$-particles, $\rho_s$ is the spectral density, which 
depends on the structure of the interactions, 
and $\mathcal{C}_0$ is the leading matching coefficient in the HQE of the differential 
width $d\Gamma/dr = \Gamma^0 \rho_s \mathcal{C}_0$.

\begin{figure}[!htb]  
	\centering
	\includegraphics[width=0.7\textwidth]{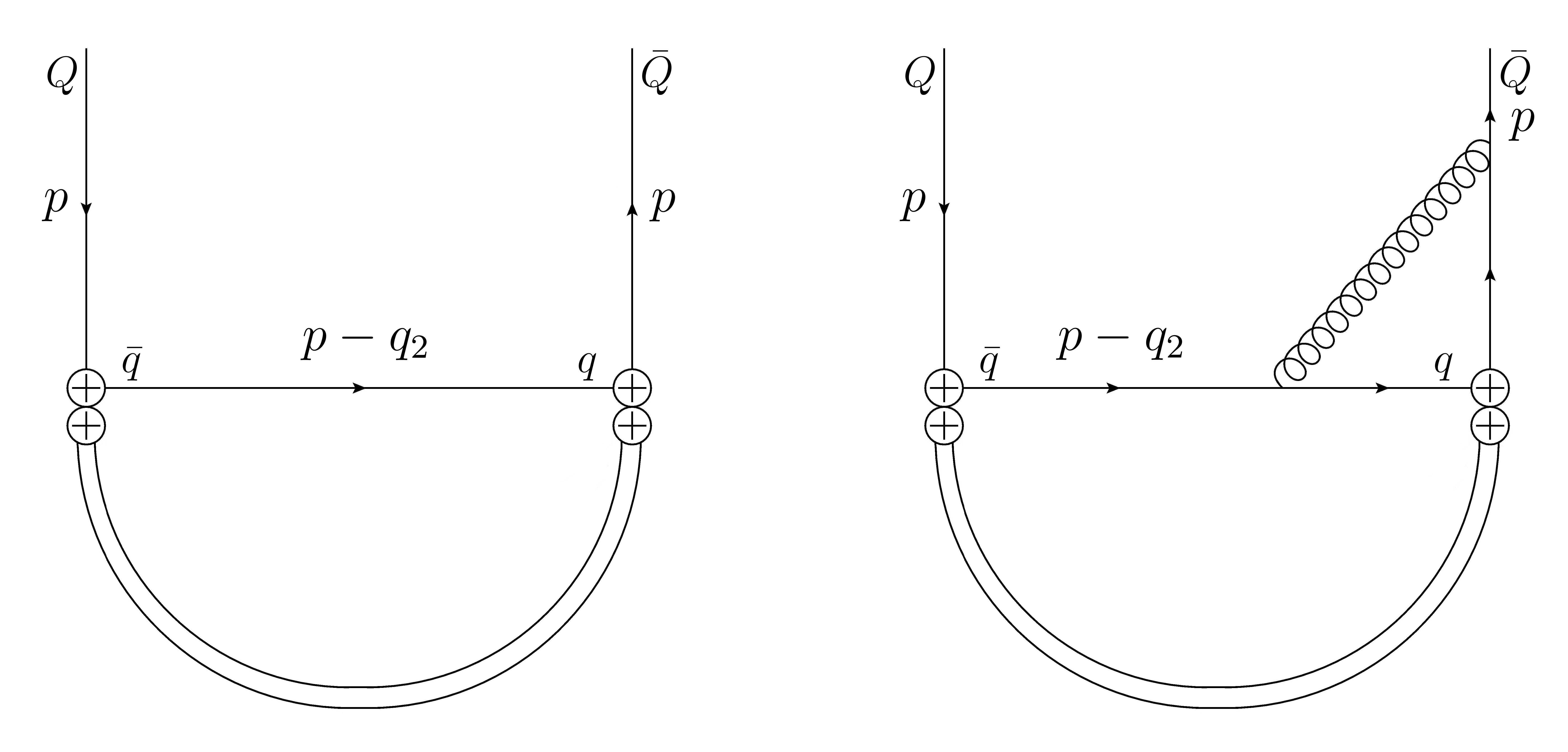}
        \caption{Examples of Feynman diagrams entering in the calculation of the coefficient of the differential width $\mathcal{C}_0$ at 
        LO-QCD (left) and NLO-QCD (right).}
        \label{GdifGEN}
\end{figure}

Note that, whereas the computation of the coefficient of the total width requires computing $n$-loop and $(n+1)$-loop diagrams with 
%$n+1$ 
scales ($\rho$, $\eta_i$) at LO-QCD and at NLO-QCD, respectively (see Fig.~[\ref{GtotGEN}]), the computation of the coefficient of 
the differential width requires computing one-loop and two-loop diagrams with two scales ($r,\rho$) at LO-QCD and at NLO-QCD, respectively 
(see Fig.~[\ref{GdifGEN}]). However, for the complete determination of the differential width one has to deal with $(n-1)$-loop integrals 
with %$n$ 
scales ($\eta_i$) to obtain the spectral density $\rho_s$.

For the computation, we use LiteRed to write the amplitudes to a combination of a small set of master integrals.
The main task of this paper, which we address in Secs. \ref{Sec:MILOQCDbclv} 
and \ref{Sec:MINLOQCDbclv}, is the analytical computation 
of the necessary master integrals for the determination of $\mathcal{C}_0(r,\rho)$ at NLO in $\alpha_s$. That requires dealing with up to 
two-loop integrals with two scales $r$ and $\rho$. These master integrals are universal to any diagrams of the type shown in Fig.~[\ref{GtotGEN}].
As a check, in Secs.~\ref{Sec:MILOQCDbulv} and \ref{Sec:MINLOQCDbulv}, we independently compute the master integrals in the case where the quark $q$ 
is massless ($m=0$) and compare to the massive case in the limit $\rho \rightarrow 0$.
From now on, we will refer to the calculation with $m\neq 0$ as the massive case and the calculation with $m=0$ as the massless case. 

The coefficient of the differential width is related to the coefficient of the total width 
through

\begin{eqnarray}
 C_0(\rho,\eta_i) &=& \int_{(\sum_{i=1}^n \sqrt{\eta_i})^2}^{(1-\sqrt{\rho})^2}dr\,\rho_s(r,\eta_i) \mathcal{C}_0(r,\rho)\,.
\end{eqnarray}
Finally, we emphasize that the differential rate is of great importance since it allows us to implement cuts on the moments 
which are unavoidable for the comparison with the experiment.
%when comparing to experiment where measures with phase space 
%cuts are unavoidable. 
The (not yet normalized) moments with a lower cut $r_{\rm min}$ and upper cut $r_{\rm max}$ are defined as 
\begin{eqnarray}
  M_n (\rho,\eta_i,r_{\mbox{\scriptsize min}},r_{\mbox{\scriptsize max}}) &=& 
  \int_{r_{\mbox{\scriptsize min}}}^{r_{\mbox{\scriptsize max}}}dr\, r^n \frac{d\Gamma(r,\rho,\eta_i)}{dr} \,,
\end{eqnarray}
with $(\sum_{i=1}^n \sqrt{\eta_i})^2<r_{\mbox{\scriptsize min}}<r_{\mbox{\scriptsize max}}<(1-\sqrt{\rho})^2$.
Obviously, having the expression for the 
differential width is more powerful than having particular moments of the distribution, since it is more general. 

%The key achievement of this paper is that the new setup is such that it contains all necessary ingredients (master integrals) for the computation of 
%$\alpha_s$ corrections to power suppressed terms ($\Lambda_{\mbox{\scriptsize QCD}}/M$ corrections).
%and the possibility to compute moments
%with cuts in the invariant mass square of the $n$ $p$-particles. 

Note that the coefficient of the differential width can  always be computed analytically. 
%Being able to get a completely analytical result or not will depend on the complexity of the spectral density $\rho_s$.
If $\rho_s$ can be computed analytically, then the differential width is known analytically and the final integration over 
$r$ to obtain the total width or moments (with or without cuts) can be
always performed, at least numerically. Numerical evaluation, if
reliable,
is practical for comparison to experimental data (see,
e.g.~\cite{Narison:1994zt} where four-loop integrals with masses have
been computed).
  %In those cases where the appearance of too many scales makes the analytical integration impossible, this representation allows to 
%write the integrals in a form ready for numerical integration.
% which is enough for comparison to experiment.

Let us stress that the master integrals computed in this paper have many physical applications. They can be used for the 
computation of the differential and total semi-leptonic and non-leptonic decay widths with any number of different masses.
%or even to topologically analogous decays with any number of massive lines. 
For semi-leptonic and non-leptonic decays the spectral density can  always be obtained analytically. Therefore, the total width and moments 
can be computed straightforwardly, at least numerically. This also applies to power corrections. 

While we will formulate the setup and the calculation in a very general language, the main motivation 
(and probably also the main application) are semileptonic decays into light massless leptons or $\tau$ leptons. 
The relevant observables are in this case the cut moments of the leptonic invariant mass spectrum, which can be used 
to extract the value of $V_{cb}$ from the data on inclusive $b \to c \ell \bar{\nu}$ transitions. In fact, 
it has been shown in \cite{Fael:2018vsp} that observables based on the leptonic invariant mass spectrum depend 
only on a reduced set of non-pertuarbative parameters and thus an even more precise determination of $V_{cb}$ 
becomes possible. 
%While we consider (in a mathematical/computational) part of the paper
%a very general formulation of the relevant correlators (and related
%masters),
%the main motivation from physics point of view is, of course, an
%analysis of the differential rate for semi-leptonic, which is a partial case of
%Eq.~(\ref{disp1LSm1m2Generic}) with two lines ($p_1$,$p_2$) which are leptons. 
%Also the most interesting case phenomenologically is when both leptons are
%massless (electron, neutrino). Even neutrino-tau decay is a bit exotic as the data
%is not precise enough at present but maybe the precision of data
%in this decay will be better soon. That will make it competitive for a
%thoughtful analysis in the nearest future.
%However, from the computational point of view we can do more than just
%semi-leptonic and therefore we present this general formulation for completeness.
%It can be used for new physics if some models fit.
%Like the ($p_1$,$p_2$) lines can be light axions or Dark matter particles
%or both, or mixture.

\section{Master integrals}
\label{masterint}

In this section, we perform the analytical computation of the master integrals necessary for the the determination of the coefficient of the 
differential width $\mathcal{C}_0$. The key ingredient is a clever choice of variables which is introduced in the next section. Based on this we will 
compute the master integrals for the massive and the massless cases.

\subsection{Choice of variables}
\label{Sec:coord}

In the massive case, computing the differential width at NLO in $\alpha_s$ requires 
the evaluation of two-loop integrals with three masses (two scales). That makes the problem 
of an analytic computation of the master integrals rather involved. However, the calculation is facilitated 
once a proper choice of the variables is made, which is motivated by the physics of the system. 
It is remarkable that such a smart choice of variables allows us to compute analytically the master integrals 
including also the full dependence on $\epsilon$, with only a single exception, where we can give an analytic 
result only as an $\epsilon$-expansion to the necessary order. However, that is sufficient for the phenomenological applications.  

Furthermore, in order to arrive at an analytical expression for the total rate, an integration over 
the $n$ $p$-particle invariant mass square has to be performed. This step is also greatly facilitated 
by a proper choice of variables. 

The naive choice would be to use $r$ and $\rho$ as the most natural variables, but when trying to compute 
two-loop sunset type integrals with two massive lines one quickly learns that this choice is not appropriate, 
since it is not adapted to the physics of the problem. In turn, the analytical dependence on $r$ and $\rho$ 
is complicated; in fact it is too complicated for any known tool to be able to perform the integration analytically. 

Rather we suggest a different set of variables ($x_{\pm}$) which is motivated by the zeros of the 
function which develops the cut and which define the integration region  
in the imaginary part of the one-loop topology with two propagators. To see this, 
it is enough to take the left hand side of 
Eq.~(\ref{disp1LSm1m2}), introduce standard Feynman parametrization, integrate over the loop-momentum and take the imaginary part. One then gets an 
integral of the form: 
\begin{equation}
 \label{howxpm}
 \mbox{Im}\,\frac{1}{i}\int \frac{d^D k}{(2\pi)^D}\frac{1}{(k^2-m^2)((p-k)^2-\chi^2)}
 \sim \int_{x_{-}}^{x_{+}} dx [-(x-x_{+})(x-x_{-})]^{-\epsilon}\,,
\end{equation}
with $D = 4 - 2 \epsilon$. This has a simple form in terms of the new variables $x_{\pm}$ which are related to 
$r$ and $\rho$ through
\begin{eqnarray}
 \label{xminus}
 x_{-} &=& \frac{1}{2}\bigg(1-r+\rho - \sqrt{(1-(\sqrt{r}-\sqrt{\rho})^2)(1-(\sqrt{r}+\sqrt{\rho})^2)}\bigg)\,,
 \\
 x_{+} &=& \frac{1}{2}\bigg(1-r+\rho + \sqrt{(1-(\sqrt{r}-\sqrt{\rho})^2)(1-(\sqrt{r}+\sqrt{\rho})^2)}\bigg)\,,
 \label{xplus}
\end{eqnarray}
Note that $x_{\pm}$ are real, $0<x_{-}<x_{+}<1$ and
\begin{eqnarray}
 x_{+} - x_{-} &=& \sqrt{(1-(\sqrt{r}-\sqrt{\rho})^2)(1-(\sqrt{r}+\sqrt{\rho})^2)}\,,
 \\
 x_{+} + x_{-} &=& 1-r+\rho\,,
\end{eqnarray}
while the Jacobian is determined from the derivatives 
\begin{eqnarray}
 \frac{dx_{+}}{dr} &=& - \frac{x_{+}}{x_{+}-x_{-}}\,,\quad\quad\quad\quad \frac{dx_{+}}{d\rho} = -\frac{1 - x_{+}}{x_{+}-x_{-}}\,,
 \\
 \frac{dx_{-}}{dr} &=& \frac{x_{-}}{x_{+} - x_{-}}\,,\quad\quad\quad\quad \frac{dx_{-}}{d\rho} = \frac{1-x_{-}}{x_{+}-x_{-}}\,.
\end{eqnarray}
It is remarkable that the inverse transformation is extremely simple
\begin{eqnarray}
 \rho &=& x_{+} x_{-}\,,
 \\
 r &=& (1-x_{+})(1-x_{-})\,,
\end{eqnarray}
compared to Eqs. (\ref{xminus}) and (\ref{xplus}). This makes the transformation of an expression 
in terms of $r$ and $\rho$ to expressions in terms of $x_{\pm}$ very easy, while the reverse is much more 
complicated. We are now ready to compute the relevant master integrals.

\subsection{LO-QCD (Massive channel)}
\label{Sec:MILOQCDbclv}

\begin{figure}[!htb]  
	\centering
	\includegraphics[width=0.9\textwidth]{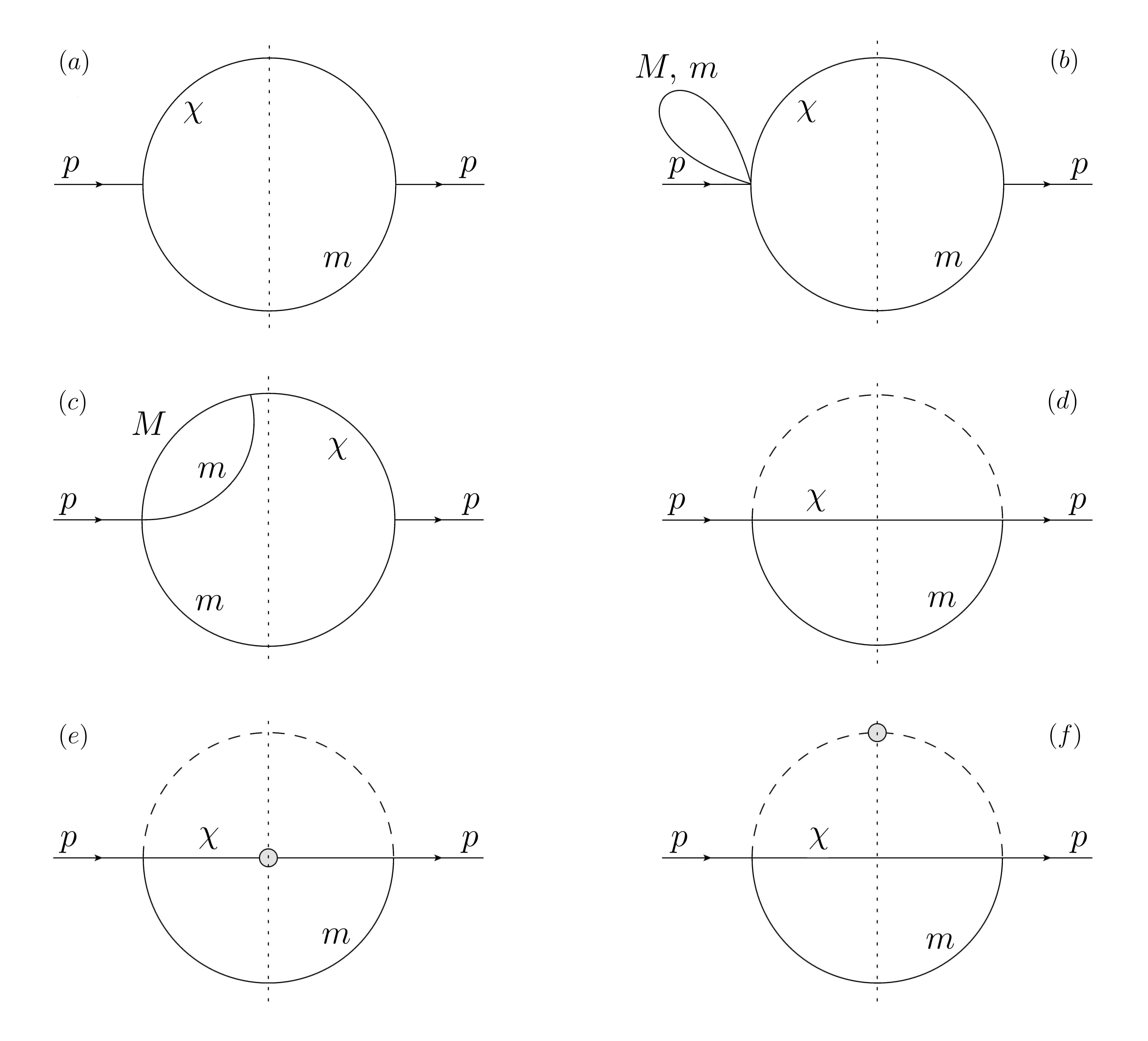}
        \caption{Master integral topologies stemming from the massive decay channel 
        contributing to the differential decay width up to NLO-QCD. Continuous and dashed lines stand for massive and massless propagators, respectively. 
        The gray dot stands for iteration of the corresponding propagator and the vertical dotted line represents a cut.}
        \label{MIsGraph}
\end{figure}

At LO-QCD the differential decay width for the massive channel 
can be written as a combination of one-loop two-propagator integrals with two different masses $m$, $\chi$
and on-shell external momenta $p^2 = M^2$ of the form
\begin{equation}
 \mathcal{M}(n_1,n_2) \equiv \mbox{Im}\, \bar\mu^{2\epsilon} \frac{1}{i}
 \int \frac{d^D q}{(2\pi)^D} \frac{1}{(q^2-\chi^2)^{n_1} ((p-q)^2 - m^2)^{n_2}}\,,
\end{equation}
where $\bar \mu^{2} = \mu^{2}(e^{\gamma_E}/4\pi)$ is the $\overline{\mbox{MS}}$ renormalization scale. 
By using LiteRed we reduce every integral above as a combination of the three master integrals $\mathcal{M}(0,1)$, $\mathcal{M}(1,0)$ 
and $\mathcal{M}(1,1)$. The master integrals $\mathcal{M}(0,1)$ and $\mathcal{M}(1,0)$ are closed massive loops and their 
imaginary part is zero. The master integral $\mathcal{M}(1,1)$ is a one-loop sunset integral with two massive lines of different 
mass, whose associated graph is displayed in Fig. [\ref{MIsGraph}] (a). It can be easily computed by using the dispersion representation given in Eq.~(\ref{disp1LSm1m2}). It reads

\begin{eqnarray}
 \mathcal{M}(1,1) &=& \left(\frac{4\pi \bar\mu^2}{M^2}\right)^{\epsilon}
 \frac{1}{16\pi}\frac{\Gamma(1-\epsilon)}{\Gamma(2-2\epsilon)}
 (x_{+} - x_{-})^{1-2\epsilon}
 \Theta(1 - \sqrt{r} - \sqrt{\rho})\,.
\end{eqnarray}
%The $\epsilon$-expansion is needed at $\mathcal{O}(\epsilon)$ since it contributes at NLO-QCD after replacing the bare bottom quark field and 
%charm quark mass by the renormalized ones at one loop order.
Finally, we have explicitly checked that our results for the master integrals reproduce the ones in the massless case, 
which are computed in Sec.~\ref{Sec:MILOQCDbulv}.

\subsection{NLO-QCD (Massive channel)}
\label{Sec:MINLOQCDbclv}
At NLO-QCD the differential decay width for the massive channel can be written as a combination of two-loop five-propagator 
integrals with three different masses $m$, $\chi$, $M$
and on-shell external momenta $p^2 =M^2$ of the form

\begin{eqnarray}
 \mathcal{J}(n_1,n_2,n_3,n_4,n_5)  
 &\equiv& \mbox{Im}\, \bar{\mu}^{4\epsilon} 
 \int \frac{d^D q_1}{(2\pi)^D}\frac{d^D q_2}{(2\pi)^D}\frac{1}{D_1^{n_1}D_2^{n_2}D_3^{n_3}D_4^{n_4}D_5^{n_5}}\,,
 \label{MostGenInt}
\end{eqnarray}
where

\begin{eqnarray}
 D_1 = q_1^2,&&\,\quad\quad\quad\quad\quad D_2 = q_2^2 - \chi^2,\,\quad\quad\quad\quad\quad D_3 = (p-q_1)^2-M^2,
 \nonumber
 \\
 &&
 D_4 = (p - q_2)^2 - m^2,\, \quad\quad\quad
 D_5 = (p-q_1-q_2)^2 - m^2\,.
\end{eqnarray}
By using LiteRed we reduce every integral above as a combination of the following thirteen master integrals

\begin{eqnarray}
 &&\mathcal{J}(0,0,0,1,1) \,,\quad\quad\quad\quad \mathcal{J}(0,1,0,1,1)\,, \nonumber
 \\
 &&\mathcal{J}(0,0,1,0,1) \,,\quad\quad\quad\quad \mathcal{J}(0,1,1,1,0) \,,\nonumber
 \\
 &&\mathcal{J}(0,0,1,1,1) \,,\quad\quad\quad\quad \mathcal{J}(0,1,1,1,1) \,,\nonumber
 \\
 &&\mathcal{J}(0,0,2,1,1) \,,\quad\quad\quad\quad \mathcal{J}(1,1,0,0,1) \,,\nonumber
 \\
 &&\mathcal{J}(0,1,0,0,1) \,,\quad\quad\quad\quad \mathcal{J}(1,2,0,0,1) \,,\nonumber
 \\
 &&\mathcal{J}(0,1,1,0,0) \,,\quad\quad\quad\quad \mathcal{J}(2,1,0,0,1) \,,\nonumber
 \\
 && \mathcal{J}(0,1,1,0,1)\,.
  \label{m7bc}
\end{eqnarray}
The integrals in the left column are either massive tadpole or two-loop sunset-type diagrams. However, taking the cut procedure 
(i.e. taking the imaginary part) these diagrams turn out to be zero. This is on the one hand due to the fact that 
tadpole diagrams have no imaginary part and, on the other hand, due to the cut procedure puts the lines on-shell, and the digrams 
vanish since no phase space is available.      
We have checked this statements also by explicit calculation. 

%
% which after taking the cut produce 
%a final state for which there is not enough phase space in the initial state for its production (in particular, one particle of mass $M$ and 
%two particles of mass $m$), so they are zero. 

The master integrals $\mathcal{J}(0,1,0,1,1)$ and $\mathcal{J}(0,1,1,1,0)$ are factorizable into a product of the cut 
one-loop two-propagator diagram with two massive lines $m$ and $\chi$, computed in Sec. \ref{Sec:MILOQCDbclv}, and  
a closed massive loop with mass $m$ and $M$, respectively. No other cut is possible. The corresponding topologies are shown in 
Fig.[\ref{MIsGraph}] (b). They read

\begin{eqnarray}
  \mathcal{J}(0,1,0,1,1) &=& \left(\frac{4\pi \bar\mu^2}{M^2}\right)^{2\epsilon} \frac{M^2}{256\pi^3}
                             \frac{\Gamma(-1+\epsilon)\Gamma(1-\epsilon)}{\Gamma(2-2\epsilon)}
                              (x_{+}x_{-})^{1-\epsilon} (x_{+} - x_{-})^{1-2\epsilon}
   \nonumber
   \\
   &&
   \times \Theta(1-\sqrt{r}-\sqrt{\rho})\,,
   \\
  \mathcal{J}(0,1,1,1,0) &=& 
   \left( \frac{4\pi\bar\mu^2}{M^2} \right)^{2\epsilon}\frac{M^2}{256\pi^3}  \frac{\Gamma(-1+\epsilon)\Gamma(1-\epsilon)}{\Gamma(2-2\epsilon)}
   (x_{+} - x_{-})^{1-2\epsilon}
   \nonumber
   \\
   &&
   \times\Theta(1-\sqrt{r}-\sqrt{\rho})\,.
\end{eqnarray}
Note that $\mathcal{J}(0,1,0,1,1)$ does not contribute to the massless case, since $x_- x_+ = \rho = 0$.

The Feynman graph representing the master integral $\mathcal{J}(0,1,1,1,1)$, which is a two-loop four-massive 
propagator integral with three different masses, is shown in Fig.[\ref{MIsGraph}] (c). 
The only way to get a non-zero imaginary part is cutting the $m$ and $\chi$ propagators. Cutting more lines 
results in a vanishing contribution by the same phase space argument discussed above.
Since only two propagators are cut, the computation is drastically simplified 
and the integral can be evaluated analytically in its full $\epsilon$ dependence. 

For the explicit computation we can first integrate the loop without cut 
while taking the "external" $\chi$-propagator momentum to be on-shell $q_2^2=\chi^2$, since we eventually will cut this line. 
The corresponding integral can be expressed in terms of hypergeometric functions of the kind $_2 F_1$. Finally, we can take the 
imaginary part of the remaining loop integral by using Eq.~(\ref{disp1LSm1m2}). The result reads

\begin{eqnarray}
 \mathcal{J}(0,1,1,1,1) &=& 
 \left( \frac{4\pi\bar\mu^2}{M^2} \right)^{2\epsilon} \frac{1}{128\pi^3} \frac{\Gamma(1-\epsilon)\Gamma(\epsilon)}
  {\Gamma(3-2\epsilon)}
  \frac{(x_{+} - x_{-})^{1-3\epsilon}(-1+x_{+})^{\epsilon}}{1-x_{+}}
  \nonumber
  \\
  &&
  \times\bigg[ 
   x_{+}^{1-\epsilon}  \, _2 F_1\bigg(1-\epsilon,\epsilon;2-\epsilon;\frac{ x_{+}(1-x_{-}) }{x_{+}-x_{-}}\bigg)
  \nonumber
  \\
  &&
  - \, _2 F_1\bigg(1-\epsilon,\epsilon;2-\epsilon;\frac{1-x_{-}}{x_{+}-x_{-}}\bigg)\bigg]
    \Theta(1-\sqrt{r}-\sqrt{\rho})\,.
\end{eqnarray}
%We have checked that the integral reproduces the massless case.
Finally, the Feynman graphs representing the master integrals $\mathcal{J}(1,1,0,0,1)$, $\mathcal{J}(1,2,0,0,1)$
and $\mathcal{J}(2,1,0,0,1)$ are shown in Fig. [\ref{MIsGraph}] (d), (e) and (f), respectively. These are two-loop integrals of the sunset 
type with one massless line and two massive lines of different mass, $\chi$ and $m$. They differ by 
the powers of the denominators. The master integrals $\mathcal{J}(1,2,0,0,1)$ and $\mathcal{J}(2,1,0,0,1)$ can be related to 
derivatives of $\mathcal{J}(1,1,0,0,1)$ with respect to the masses

\begin{eqnarray}
 \mathcal{J}(1,2,0,0,1) &=& \frac{d}{d\chi^2}\mathcal{J}(1,1,0,0,1) \,,
 \\
 \label{j21001bcrelder}
 \mathcal{J}(2,1,0,0,1) &=& \frac{1}{M^4 \epsilon(r^2 + (-1+\rho)^2 - 2r(1+\rho))}
 \bigg[2(-1+\epsilon)^2 \mathcal{J}(0,1,0,0,1) 
 \\
 &&
 - M^2(-1+2\epsilon)\bigg((-2+3\epsilon)(-1+r+\rho)\mathcal{J}(1,1,0,0,1)
 \nonumber
 \\
 &&
 + 2M^2 \Big((-1+\rho)\rho \frac{d}{dm^2}\mathcal{J}(1,1,0,0,1) + (-1+r)r \frac{d}{d\chi^2}\mathcal{J}(1,1,0,0,1)\Big)\bigg)\bigg]\,,
 \nonumber
\end{eqnarray}
where Eq.~(\ref{j21001bcrelder}) have been obtained by writing $\mathcal{J}(1,1,0,0,2)$ as a combination of the master 
integrals in Eq.~(\ref{m7bc})
by using LiteRed and solving for $\mathcal{J}(2,1,0,0,1)$. Alternatively, in terms of $x_{\pm}$, the relations to $\mathcal{J}(1,1,0,0,1)$ 
read

\begin{eqnarray}
\mathcal{J}(1,2,0,0,1) &=&
 \frac{1}{M^2}\frac{1}{x_{+}-x_{-}}
 \bigg(x_{-}\frac{\partial}{\partial x_{-}} - x_{+} \frac{\partial}{\partial x_{+}}  \bigg)\mathcal{J}(1,1,0,0,1)\,,
 \label{j12001relderxpxm}
 \\
 \mathcal{J}(2,1,0,0,1) &=& 
 \frac{2(1-2\epsilon)}{M^2 \epsilon (x_{+}-x_{-})^2}
 \bigg[
  \frac{1}{2}(2-3\epsilon)(x_{+} + x_{-} - 2x_{+}x_{-})\mathcal{J}(1,1,0,0,1)
 \nonumber
 \\
 &&
 + \frac{x_{+}(1 - x_{+})}{x_{+}-x_{-}} \bigg( 
   (1 - x_{+}x_{-}) x_{-}
 + (x_{+} + x_{-} - x_{+}x_{-})(1-x_{-})
 \bigg)
 \nonumber
 \\
 &&
 \quad\quad\times
 \frac{\partial}{\partial x_{+}} \mathcal{J}(1,1,0,0,1)
 \nonumber
 \\
 &&
 - \frac{x_{-}(1-x_{-})}{x_{+}-x_{-}} \bigg( (1 - x_{+}x_{-}) x_{+}
 + (x_{+} + x_{-} - x_{+}x_{-})(1-x_{+})
 \bigg)
 \nonumber
 \\
 &&
 \quad\quad\times
 \frac{\partial}{\partial x_{-}} \mathcal{J}(1,1,0,0,1)
 \bigg]\,.
 \label{j21001relderxpxm}
\end{eqnarray}
Therefore, the main problem is the computation of 
$\mathcal{J}(1,1,0,0,1)$. This integral is the most complicated one, since it requires cutting one massless line and 
two massive lines of different mass. In this particular case, making use of $x_{\pm}$ instead of $r$ and $\rho$ 
makes a crucial difference, allowing us to compute this contribution also analytically, however, only in an 
$\epsilon$-expansion.

For the computation we first introduce Feynman parametrization in the massive loop and integrate over the loop momenta. Next we make use 
of the generalized spectral function with one massive line~\cite{Mannel:2014xza,Mannel:2015jka}

\begin{eqnarray}
 \mbox{Im}\, \frac{1}{i}\int \frac{d^D k}{(m_0^2 - k^2)^a (-(p-k)^2)^b}
  &=&  \frac{\pi^{1+D/2}\Gamma(D/2-b)}{\Gamma(a)\Gamma(b)\Gamma(D-a-2b+1)}
 \bigg(1-\frac{m_0^2}{M^2}\bigg)^{D-a-2b} M^{D-2a-2b}
 \nonumber
 \\
 &&
 \times
 \,_2 F_1\bigg(D/2-b,1-b;D-a-2b+1;1-\frac{m_0^2}{M^2}\bigg)
\end{eqnarray}
to compute the imaginary part, where in our case

\begin{equation}
 m_0^2 = \frac{m^2}{1-x} + \frac{\chi^2}{x}
\end{equation}
is a function of $m$, $\chi$ and a single Feynman parameter $x$. The branch cut appears always that 
$M^2>m_0^2(m,\chi,x)$, which constrains the integration limits of the Feynman parameter integral from $x_{-}$ to $x_{+}$, instead 
of from zero to one. 
%That makes clear again why it is better to choose as natural variables $x_{\pm}$ instead of $r$ and $\rho$. 
Finally, we end up with an integral of the form
\begin{eqnarray}
 \mathcal{J}(1,1,0,0,1) &\propto&  \int_{x_{-}}^{x_{+}} dx  (x(1-x))^{-2+2\epsilon} \bigg( -(x-x_{+})(x-x_{-}) \bigg)^{2-3\epsilon}\,,
\end{eqnarray}
which can not be expressed in terms of the usual $_2 F_1$-hypergeometric functions. 
It requires further study if it can be expressed in terms of 
generalized  $_p F_q$-hypergeometric functions or if it requires the introduction of the 
class of functions called elliptic polylogarithms, which 
are known to appear in two-loop sunset diagrams with three massive lines~\cite{Adams:2016vdo}. 

However, the integral can be computed as 
an $\epsilon$ expansion to the necessary order,  the reason is related to the fact that one 
of the lines is massless. 
In general we need the master integrals expanded to $\mathcal{O}(\epsilon)$, 
because the coefficients in front of the master integrals can be proportional to $1/\epsilon$. 

We first note that the integral is finite. 
The $\mathcal{O}(\epsilon^0)$ term is trivial since it is the integral of a rational function and 
can be directly computed. 
The $\mathcal{O}(\epsilon)$ term is an integral of the form 
$R(x)\ln(P(x))$, where $R(x)$ is a rational function and $P(x)$ a polynomial, 
and thus the integral can be expressed in terms of dilogarithms~\cite{Kirillov:1994en}.
%Mathematica has many difficulties to recognize the integral 
%and computational time may last to infinity if we try to perform the integration blindly, so it needs extra help. We expand $R(x)$ in simpler 
%factors using the Mathematica command Apart[]
%and integrate term by term. 
The resulting expression reads

\begin{eqnarray}
 &&\mathcal{J}(1,1,0,0,1) = 
 \nonumber
 \\
 &&
   \frac{M^2}{512\pi^3}\bigg\{  \bigg[1 + \frac{\epsilon}{2}\bigg(9 + 8 \ln \bigg(\frac{\mu }{M}\bigg) \bigg) \bigg]\bigg[
    (x_{+}-x_{-}) (2-x_{+}-x_{-} +2x_{+} x_{-}) 
   \nonumber
   \\
   &&
   - 2 x_{-} x_{+} (x_{+} + x_{-} - x_{+} x_{-} ) \ln\bigg(\frac{x_{+}}{x_{-}}\bigg)
   - 2 (1-x_{+})(1-x_{-})(1-x_{-} x_{+}) \ln \bigg(\frac{1-x_{-}}{1-x_{+}}\bigg) \bigg]
   \nonumber
   \\
   &&
   + \epsilon \bigg[
   2(x_{+}-x_{-})( 2 - x_{-} - x_{+} + 2 x_{-} x_{+}  )\bigg( 
   1 - 3 \ln (x_{+}-x_{-})
   \bigg)
   \nonumber
   \\
   &&
   + (1-x_{+})( x_{+}x_{-}(1 -  4x_{-}) - 4 + x_{+} + x_{-} + x_{-}^2 ) \ln (1-x_{+})
   \nonumber
   \\
   &&
   - (1-x_{-})( x_{+}x_{-}(1 - 4x_{+}) - 4 + x_{+} + x_{-} + x_{+}^2 ) \ln (1-x_{-}) 
   \nonumber
   \\
   &&
   - x_{-} ( x_{+} x_{-} (7-4 x_{+}) + 4 - 2x_{-} - 4x_{+} + 3x_{+}^2) \ln (x_{-}) 
   \nonumber
   \\
   &&
   + x_{+} ( x_{+} x_{-}(7 - 4x_{-})  + 4 - 2x_{+} - 4x_{-} + 3x_{-}^2 ) \ln (x_{+})
   \nonumber
   \\
   &&
   + 4 \bigg( - 2x_{+}x_{-}( x_{-} + x_{+} - x_{+} x_{-}) + x_{-}+x_{+}-1 \bigg) \bigg(\Li_2(x_{-})-\Li_2(x_{+})\bigg)
   \nonumber
   \\
   &&
   - 2 (1-x_{-}) (1-x_{+})(1 - x_{+} x_{-})\bigg( 
   3\Li_2\bigg(\frac{x_{-}-x_{+}}{x_{-}-1}\bigg)-3\Li_2\bigg(\frac{x_{+}-x_{-}}{x_{+}-1}\bigg) 
   + \ln ^2(1-x_{-})
   \nonumber
   \\
   &&
   -\ln^2(1-x_{+}) 
   + 2\ln (1-x_{-}) \ln (x_{-}) - 2\ln(1-x_{+}) \ln (x_{+})
   + 6\ln \bigg(\frac{1-x_{+}}{1-x_{-}}\bigg) \ln (x_{+}-x_{-})\bigg)
   \nonumber
   \\
   &&
   - 2 x_{+} x_{-} ( x_{+} + x_{-} - x_{+}x_{-}) 
   \bigg( 3\Li_2\bigg(1-\frac{x_{-}}{x_{+}}\bigg)-3\Li_2\bigg(1-\frac{x_{+}}{x_{-}}\bigg) -\ln ^2(x_{-}) + \ln^2(x_{+}) 
   \nonumber
   \\
   &&
   + 6\ln \bigg(\frac{x_{-}}{x_{+}}\bigg) \ln (x_{+}-x_{-}) \bigg)
   \bigg]
   + \mathcal{O}(\epsilon^2)\bigg\}\Theta(1-\sqrt{r}-\sqrt{\rho})\,.
\end{eqnarray}
Once $\mathcal{J}(1,1,0,0,1)$ is known we can compute $\mathcal{J}(1,2,0,0,1)$ and $\mathcal{J}(2,1,0,0,1)$ by using 
Eqs. (\ref{j12001relderxpxm}) and (\ref{j21001relderxpxm}). Explicitly, they read

\begin{eqnarray}
 \mathcal{J}(1,2,0,0,1) &=&
  \frac{1}{256 \pi^3}\bigg\{ 
 \bigg[1 + 4\epsilon \ln \bigg(\frac{\mu }{M}\bigg)\bigg]\bigg[ x_{+} - x_{-} + x_{+} x_{-} \ln \bigg(\frac{x_{+}}{x_{-}}\bigg) 
   \\
   &&
   + (1-x_{+} x_{-}) \ln \bigg(\frac{1-x_{+}}{1-x_{-}}\bigg) \bigg]
   + \epsilon \bigg[  (x_{+}-x_{-})\bigg(5-6\ln(x_{+}-x_{-}) \bigg)
    \nonumber
   \\
   &&
   -  (1+x_{-}) (1-x_{+}) \ln (1-x_{-}) 
   +  (1-x_{-}) (1+x_{+}) \ln (1-x_{+})
   \nonumber
   \\
   &&
   - x_{-} (x_{+}+2) \ln (x_{-}) 
   + x_{+} (x_{-}+2) \ln (x_{+})   
   \nonumber
   \\
   &&
   - 2 (1-2 x_{+} x_{-} )\bigg(\Li_2(x_{-})-\Li_2(x_{+})\bigg)  
   \nonumber
   \\
   &&
   + x_{+} x_{-} \bigg( 3\Li_2\bigg(1-\frac{x_{-}}{x_{+}}\bigg) - 3\Li_2\bigg(1-\frac{x_{+}}{x_{-}}\bigg) + \ln ^2(x_{+})-\ln ^2(x_{-}) 
   \nonumber
   \\
   &&
   + 6\ln \bigg(\frac{x_{-}}{x_{+}}\bigg) \ln(x_{+}-x_{-})\bigg)     
   \nonumber
   \\
   &&
   -  (1-x_{+} x_{-}) \bigg( 3\Li_2\bigg(\frac{x_{-}-x_{+}}{x_{-}-1}\bigg) - 3\Li_2\bigg(\frac{x_{+}-x_{-}}{x_{+}-1}\bigg)
   \nonumber
   \\
   &&
   + 6\ln \bigg(\frac{1-x_{+}}{1-x_{-}}\bigg) \ln (x_{+}-x_{-}) 
   + \ln^2(1-x_{-}) - \ln^2(1-x_{+})
   \nonumber
   \\
   &&
   + 2\ln (1-x_{-}) \ln (x_{-}) -2\ln (1-x_{+}) \ln (x_{+}) 
   \bigg)
   \bigg]
   + \mathcal{O}(\epsilon^2)\bigg\}\Theta(1-\sqrt{r}-\sqrt{\rho})\,,
   \nonumber
   \\
    \mathcal{J}(2,1,0,0,1) &=& 
  \frac{1}{256 \pi ^3}\bigg\{ (x_{+}-x_{-})\bigg[\frac{1}{\epsilon}
   + 4 \ln \bigg(\frac{\mu }{M}\bigg)
   - 6 \ln (x_{+}-x_{-}) 
   + 3\bigg]
   \\
   &&
   + (1-x_{-})(1+x_{+}) \ln (1-x_{-})
   - (1-x_{+})(1+x_{-}) \ln (1-x_{+}) 
   \nonumber
   \\
   &&
   + x_{-} (x_{+}-2) \ln (x_{-})
   - x_{+} (x_{-}-2) \ln (x_{+})
   + \mathcal{O}(\epsilon)
   \bigg\}\Theta(1-\sqrt{r}-\sqrt{\rho})\,.
   \nonumber
\end{eqnarray}
Note that $\mathcal{J}(2,1,0,0,1)\sim (1/\epsilon)\mathcal{J}(1,1,0,0,1)$. Therefore, in order to compute 
$\mathcal{J}(2,1,0,0,1)$ to $\mathcal{O}(\epsilon)$ we need $\mathcal{J}(1,1,0,0,1)$ to $\mathcal{O}(\epsilon^2)$, 
whose computation can be rather cumbersome but possible. Due to the length of the resulting expression we do not present it explicitly 
but we provide it upon request. 

However, for the particular applications we will discuss in Sec.~\ref{Sec:applications} the coefficient in front of 
$\mathcal{J}(2,1,0,0,1)$ coming from the IBP reduction is $\mathcal{O}(\epsilon^0)$, so the integral is only needed to $\mathcal{O}(\epsilon^0)$, and thus knowing 
$\mathcal{J}(1,1,0,0,1)$ to $\mathcal{O}(\epsilon)$ is actually enough.

Note that some of the master integrals are divergent as $\epsilon \to 0$, so in principle there can even 
be $1/\epsilon^2$ terms. However, in the expressions for physical quantities such as the differential widths 
the master integrals must combine in such a way that 
the $1/\epsilon^2$ term cancels. This is to be expected since after taking cut we face one-loop integrals, so $1/\epsilon^2$ terms 
should never appear. A different way of seeing this is that at $\mathcal{O}(\alpha_s)$ poles must cancel with the one-loop renormalization 
factors of the fields and masses which, to this order, contain a simple pole in $\epsilon$ at most. 
The explicit cancellation of $1/\epsilon^2$ terms in the applications discussed in Sec.~\ref{Sec:applications} is thus 
a check of our calculation. 

Let us mention that there is a peculiarity related to the $1/\epsilon^2$ terms. 
In a one-loop calculation it is usually enough to take only the finite piece which involves only logarithms. However since our original 
integrals contain $1/\epsilon^2$, which is a two-loop feature, the finite piece contains also dilogarithms and factors of $\pi^2$. 
Even though in general the $1/\epsilon^2$ cancels in the width, a combination of dilogarithms and factors of $\pi^2$ survive. 
These surviving terms are a remnant of the extra divergences which appeared in intermediate steps. 
This extra divergences are related to the fact that the $q$ quark 
is on-shell. A similar mechanism is at work in the matching of heavy-light currents in SCET and collinear divergences~\cite{Bell:2010mg}.

Finally, we have explicitly checked that our results for the master integrals reproduce the ones in the massless case, 
which are computed below  in Sec.~\ref{Sec:MINLOQCDbulv}.

\subsection{LO-QCD (Massless channel)}
\label{Sec:MILOQCDbulv}

\begin{figure}[!htb]  
	\centering
	\includegraphics[width=0.9\textwidth]{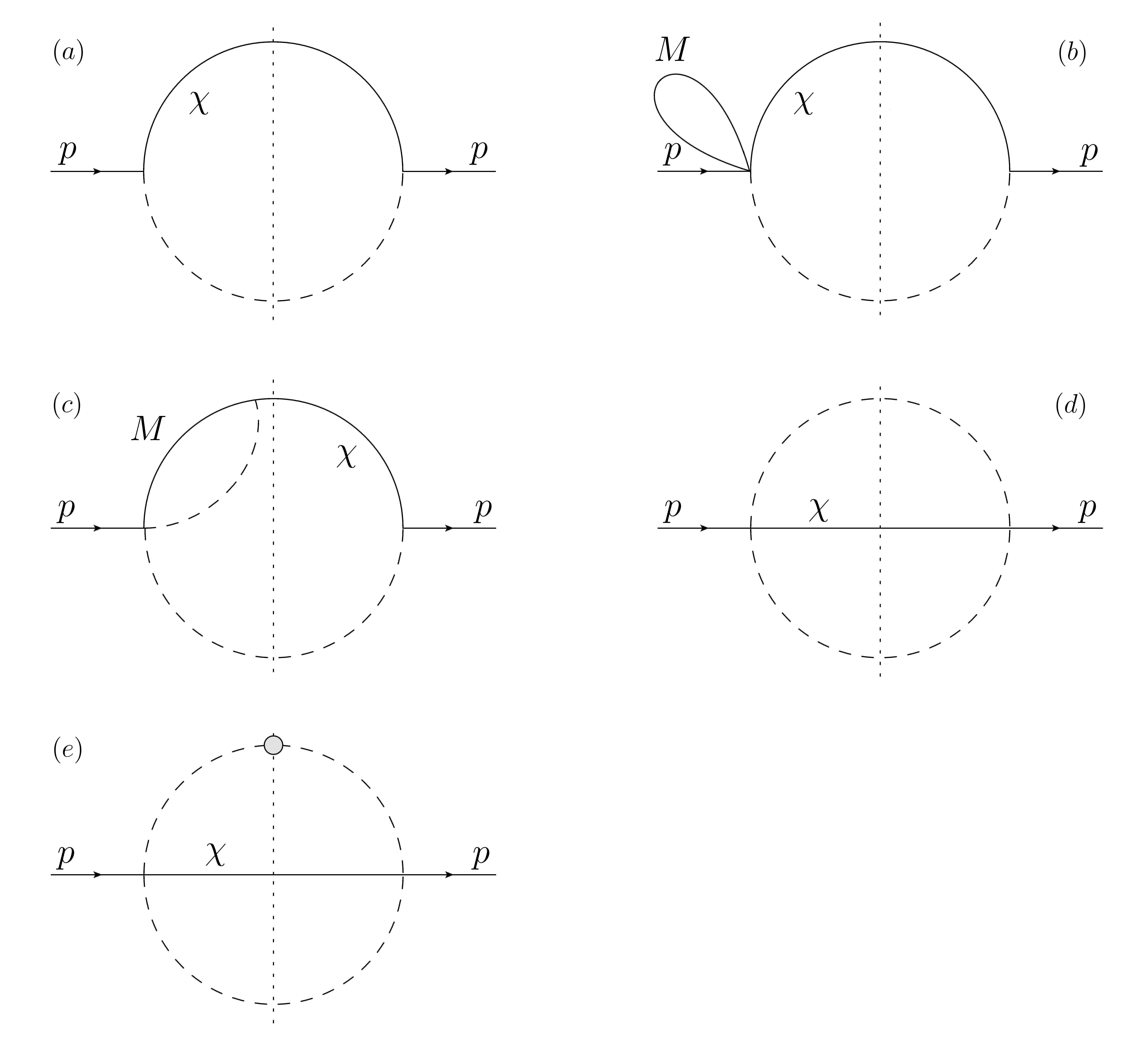}
        \caption{Master integral topologies stemming from the massless decay channel 
        contributing to the differential decay width up to NLO-QCD. Continuous and dashed lines stand for massive and massless propagators, respectively. 
        The gray dot stands for iteration of the corresponding propagator and the vertical dotted line represents a cut.}
        \label{MIm0Graph}
\end{figure}

At LO-QCD the differential decay width for the massless ($\rho=0$) channel can be written as a combination of 
one-loop two-propagator integrals with a single mass $\chi$
and on-shell external momenta $p^2 =M^2$ of the form
\begin{eqnarray}
 \mathcal{M}(n_1,n_2) &\equiv& \mbox{Im}\, \bar{\mu}^{2\epsilon}
  \frac{1}{i}\int \frac{d^D q}{(2\pi)^D} \frac{1}{(q^2-\chi^2)^{n_1} ((p-q)^2)^{n_2}}\,.
\end{eqnarray}
By using LiteRed we reduce every integral above as a combination of the two master integrals $\mathcal{M}(1,0)$ and 
$\mathcal{M}(1,1)$. The master integral $\mathcal{M}(1,0)$ is a closed massive loop and its 
imaginary part is zero. The master integral $\mathcal{M}(1,1)$ is a one-loop integral with one massive and one massless line, 
whose associated graph is 
displayed in Fig.~[\ref{MIm0Graph}] (a).
It can be easily computed by using the dispersion representation given in Eq.~(\ref{disp1LSm1m2}). It reads

\begin{eqnarray}
 \mathcal{M}(1,1) &=& \left(\frac{4\pi\bar{\mu}^2}{M^2} \right)^{\epsilon} \frac{1}{16\pi}
 \frac{\Gamma(1-\epsilon)}{\Gamma(2-2\epsilon)}(1-r)^{1-2\epsilon}\Theta(1-r)\,.
\end{eqnarray}
%The $\epsilon$-expansion is needed at $\mathcal{O}(\epsilon)$ since it contributes at NLO-QCD after replacing the bare bottom quark field 
%by the renormalized one at one loop order.

\subsection{NLO-QCD (Massless channel)}
\label{Sec:MINLOQCDbulv}

At NLO-QCD the differential decay width for the massless ($\rho=0$) channel can be written as a combination of two-loop five-propagator 
integrals with two different masses $\chi$ and $M$ and on-shell external momenta $p^2 =M^2$ of the form

\begin{eqnarray}
 \mathcal{J}(n_1,n_2,n_3,n_4,n_5)  
 &\equiv& \mbox{Im}\, \bar{\mu}^{4\epsilon}\int \frac{d^D q_1}{(2\pi)^D}\frac{d^D q_2}{(2\pi)^D}\frac{1}{D_1^{n_1}D_2^{n_2}D_3^{n_3}D_4^{n_4}D_5^{n_5}}\,,
 \label{MostGenInt}
\end{eqnarray}
where

\begin{eqnarray}
 D_1 = q_1^2,&&\,\quad\quad\quad\quad\quad D_2 = q_2^2 - \chi^2,\,\quad\quad\quad\quad\quad D_3 = (p-q_1)^2-M^2,
 \nonumber
 \\
 &&
 D_4 = (p - q_2)^2 ,\, \quad\quad\quad
 D_5 = (p-q_1-q_2)^2 \,.
\end{eqnarray}
By using LiteRed we reduce every integral above as a combination of the following seven master integrals

\begin{eqnarray}
 && \mathcal{J}(0,1,1,1,0) \,,\quad\quad\quad\quad \mathcal{J}(0,0,1,1,1) \,,\quad\quad\quad\quad
 \\
 && \mathcal{J}(0,1,1,1,1) \,,\quad\quad\quad\quad \mathcal{J}(0,1,1,0,0) \,,\quad\quad\quad\quad
 \\
 && \mathcal{J}(1,1,0,0,1) \,,\quad\quad\quad\quad \mathcal{J}(0,1,1,0,1) \,,\quad\quad\quad\quad
 \\
 && \mathcal{J}(2,1,0,0,1)\,.
\end{eqnarray}
The master integrals in the right column are either closed loops or two-loop sunset type diagrams which - after taking the cut - vanish by 
the same arguments as for the massive case. They are either massive tadpoles (which are real) or there
is no phase space. Again, we have checked this statement by explicit calculation as well.

The master integral $\mathcal{J}(0,1,1,1,0)$ is factorizable in the product of the cut 
one-loop sunset integral with one massive line $\chi$, computed in Sec. \ref{Sec:MILOQCDbulv}, times 
a closed massive loop with mass $M$. Any other cut is not possible. The corresponding topology is shown in 
Fig.~[\ref{MIm0Graph}] (b). The result reads

\begin{eqnarray}
 \mathcal{J}(0, 1, 1, 1, 0) &=& \left(\frac{4\pi\bar{\mu}^2}{M^2}\right)^{2\epsilon} \frac{M^2}{256\pi^3} 
                                \frac{\Gamma(-1+\epsilon)\Gamma(1-\epsilon)}{\Gamma(2-2\epsilon)}(1-r)^{1-2\epsilon}\Theta(1-r)\,.
\end{eqnarray}
The Feynman graph representing the master integral $\mathcal{J}(0,1,1,1,1)$, which is a two-loop two-massless and two-massive 
propagator integral with two different masses, is shown in Fig.~[\ref{MIm0Graph}] (c). For the computation we can proceed exactly
as in the completely massive case.
The only way to get a non-zero imaginary part is cutting the massless and $\chi$ propagators. 
Otherwise, there is not enough energy in the initial state to produce the final state particles 
(one particle of mass $M$ and two massless quarks). Again, the cut involves only two propagators. 
Since the loop involving the particle of mass $M$ and one massless particle is not cut we can first integrate it while taking 
the "external" $\chi$-propagator momentum to be on-shell due to 
the cut on it. Finally, we can take the imaginary part of the remaining loop 
integral by using Eq.~(\ref{disp1LSm1m2}). The result reads

\begin{eqnarray}
 \mathcal{J}(0, 1, 1, 1, 1) &=& - \left(\frac{4\pi\bar{\mu}^2}{M^2}\right)^{2\epsilon} \frac{1}{128\pi^3} 
                                       \frac{\Gamma(1-\epsilon)\Gamma(\epsilon) }{\Gamma(3-2\epsilon)}
                                       (1-r)^{1-2\epsilon} \,_2 F_1(1,\epsilon;2-\epsilon;r)
                                       \nonumber
                                       \\
                                       &&
                                       \times
                                       \Theta(1-r)\,.
\end{eqnarray}
Finally, the Feynman graphs representing the master integrals $\mathcal{J}(1,1,0,0,1)$ and 
$\mathcal{J}(2,1,0,0,1)$ are shown in Fig.~[\ref{MIm0Graph}] (d) and (e). These are two-loop integrals of the sunset type with two massless lines 
and one massive line of mass $\chi$. They differ by 
the powers of the denominators. The master integral $\mathcal{J}(2,1,0,0,1)$ can be related to 
$\mathcal{J}(1,1,0,0,1)$ and derivatives of it with respect to the mass $\chi$. Explicitly

\begin{eqnarray}
 \mathcal{J}(2, 1, 0, 0, 1)  &=& \frac{(-3+D)}{(-4+D)(M^2 - \chi^2)}\bigg[ (8-3D)\mathcal{J}(1,1,0,0,1) 
                                       + 4\chi^2 \frac{d}{d\chi^2}\mathcal{J}(1,1,0,0,1)\bigg]\,.
                                       \nonumber
                                       \\
                                       \label{j21001relbu}
\end{eqnarray}
To obtain the relation above we reduce the integral $\mathcal{J}(1,2,0,0,1)=\frac{d}{d\chi^2}\mathcal{J}(1,1,0,0,1)$ using LiteRed and invert 
the relation to find $\mathcal{J}(2,1,0,0,1)$ in terms of $\mathcal{J}(1,2,0,0,1)$ and $\mathcal{J}(1,1,0,0,1)$. 

Therefore, the problem is reduced to the computation of $\mathcal{J}(1,1,0,0,1)$. 
It requires taking a cut across two massless lines and 
one massive line. For the computation we can proceed analogously to the massive case. Since there is only one massive line 
the integration can be performed to all orders in $\epsilon$, unlike in the massive case. The result reads

\begin{eqnarray}
 \mathcal{J}(1, 1, 0, 0, 1) &=& \left(\frac{4\pi\bar{\mu}^2}{M^2}\right)^{2\epsilon}\frac{M^2}{256\pi^3} \frac{\Gamma^2(1-\epsilon)}{\Gamma(4-4\epsilon)}
 \nonumber
 \\
 &&
 \times (1-r)^{3-4\epsilon}\,_2 F_1(2-2\epsilon,1-\epsilon;4-4\epsilon;1-r)\Theta(1-r)\,.
\end{eqnarray}
Once $\mathcal{J}(1,1,0,0,1)$ is known, we can compute $\mathcal{J}(2,1,0,0,1)$ by using 
Eq.~(\ref{j21001relbu}). It reads

\begin{eqnarray}
 \mathcal{J}(2, 1, 0, 0, 1) 
                       &=& \left(\frac{4\pi\bar{\mu}^2}{M^2}\right)^{2\epsilon} \frac{1}{256\pi^3}
                       \frac{\Gamma(2\epsilon)\Gamma(2-\epsilon)
                       \Gamma(-1+\epsilon)\Gamma(1-\epsilon)}{\Gamma(1+\epsilon) \Gamma(4-4\epsilon)\Gamma(-1+2\epsilon)}
 \nonumber
 \\                      
                       &&\times (1-r)^{1-4\epsilon} \bigg[
                       (2-3\epsilon + r(4-5\epsilon))\, _2 F_1(2-2\epsilon,1-\epsilon;4-4\epsilon;1-r) 
 \nonumber
 \\                     
                       && + (1-\epsilon) r(1-r) \,_2 F_1(3-2\epsilon,2-\epsilon;5-4\epsilon;1-r)\bigg]\Theta(1-r)\,.
 \end{eqnarray}
Unlike in the massive channel case, we are able to get all master integrals in their full $\epsilon$ dependence.

\section{Applications}
\label{Sec:applications}

As a sample application, we compute the differential and total $B$-hadron ($Q= b$, $M= m_b$) semi-leptonic decay width to leading order in 
the $1/m_b$ expansion. There are two decay channels that contribute. On the one hand, there is a 
Cabibbo favoured transition $b\rightarrow c\ell\bar{\nu}_\ell$, where we identify $q= c$ and $m=m_c$.
On the other hand, there is a Cabibbo suppressed transition $b\rightarrow u \ell \bar \nu_\ell$, where we identify 
$q=u$ and $m=0$. They correspond to the massive and massless cases respectively, which have been 
discussed in Secs.~\ref{Sec:formulation} and \ref{masterint}.
In both channels $n=2$ with $p_1 =\ell$ and $p_2=\bar\nu_\ell$ being massless leptons, $m_1=m_2=0$. Again, we use the massless 
case as a check of the computation for the massive case in the limit $\rho\rightarrow 0$.

The effective electroweak Lagrangian describing the flavor changing transition $b\rightarrow q \ell\bar{\nu}_\ell$ reads~\cite{Buchalla:1995vs}

\begin{eqnarray}
 \mathcal{L}_{{\scriptsize\mbox{EW, eff}}}  &=& -\frac{4G_F}{\sqrt{2}} \sum_{q=c,u}\bigg( 
 V_{qb} (\bar b \Gamma_\mu q)(\bar \nu_\ell \Gamma^\mu \ell) 
+V_{qb}^\dagger (\bar q \Gamma_\mu b)(\bar\ell \Gamma^\mu \nu_\ell)
\bigg)\,,
\end{eqnarray}
where $G_F$ is the Fermi constant, $V_{qQ}$ is the corresponding 
CKM matrix element describing weak mixing of quark generations, 
and $\Gamma^\mu = \gamma^\mu P_L$ with $P_L=\frac{1}{2}(1-\gamma_5)$ being the left-handed projector. In the cases under study 
$\Gamma^0 = G_F^2 m_b^5 |V_{qb}|^2/(192\pi^3)$.

Before going to the technical details we need to discuss how renormalization will be performed. 
For light-quark masses and the strong coupling $\alpha_s(\mu)$ we adopt the $\overline{\mbox{MS}}$ renormalization scheme~\cite{Grozin:2005yg}.
The bottom and charm quarks will be renormalized on-shell. In practice, that is $b_B = (Z_2^{\mbox{\scriptsize OS}})^{1/2} b$ and 
$m_{c,B} = Z_{m_c}^{\mbox{\scriptsize OS}} m_c^{\mbox{\scriptsize pole}}$, where quantities with omitted indices stand for renormalized ones, and

\begin{eqnarray}
 Z_{m_q}^{\mbox{\scriptsize OS}} &=& 1 - C_F \frac{\alpha_s(\mu)}{4\pi}\bigg( \frac{3}{\epsilon} + 6 \ln\left(\frac{\mu}{m_q}\right) + 4 \bigg)\,,
%Z_2^{\mbox{\scriptsize OS}}  &=& 1 - C_F \frac{\alpha_s(\mu)}{4\pi}\bigg( \frac{3}{\epsilon} + 6 \ln\left(\frac{\mu}{m_b}\right) + 4 \bigg)\,,
% \\
% &=& 1 - C_F \frac{g_s^2}{16\pi^2}\mu^{-2\epsilon}\bigg(\frac{e^{\gamma_E}}{4\pi}\bigg)^{-\epsilon}\bigg( \frac{3}{\epsilon} + 6 \ln\left(\frac{\mu}{m_b}\right) + 4 \bigg)
% \\
% Z_{m_c}^{\mbox{\scriptsize OS}} &=& 1 - C_F \frac{\alpha_s(\mu)}{4\pi}\bigg( \frac{3}{\epsilon} + 6 \ln\left(\frac{\mu}{m_c}\right) + 4 \bigg)\,,
% \\
% &=& 1 - C_F \frac{g_s^2}{16\pi^2}\mu^{-2\epsilon}\bigg(\frac{e^{\gamma_E}}{4\pi}\bigg)^{-\epsilon}\bigg( \frac{3}{\epsilon} + 6 \ln\left(\frac{\mu}{m_c}\right) + 4 \bigg)
% \\
% &=& 1 - C_F \frac{\alpha_s}{4\pi}\bigg( \frac{3}{\epsilon} + 6 \ln\left(\frac{\mu}{m_b}\right)  - 3 \ln\left(x_{+}\right) - 3 \ln\left(x_{-}\right) + 4 \bigg)
\end{eqnarray}
with $Z_2^{\mbox{\scriptsize OS}}= Z_{m_b}^{\mbox{\scriptsize OS}}$ to this order. 
Also $g_{s,B}^2 = 4\pi Z_g^2 \alpha_s(\mu)\bar{\mu}^{2\epsilon}$. 
For the precision of the calculation the renormalization factor of the strong coupling  is only needed at tree level ($Z_g=1$). 
The quantity $C_F=4/3$ is a color factor.

Note that for the charm quark it is still not so clear if it is better to perform the renormalization in the on-shell or the $\overline{\mbox{MS}}$ scheme. 
We have chosen the former because results become slightly more compact in that scheme. However, since $m_c$ is better known in 
the $\overline{\mbox{MS}}$ scheme it might be useful to have the results with $m_c$ in that scheme. This can be easily achieved by using the 
relation between the $\overline{\mbox{MS}}$ and pole masses at one-loop order

\begin{eqnarray}
 m_c^{\mbox{\scriptsize pole}} &=& 
 m_c^{\overline{\mbox{\scriptsize MS}}}(\mu)\bigg(1 + C_F \frac{\alpha_s}{4\pi}\bigg( 6\ln\left(\frac{\mu}{m_c}\right) + 4\bigg)\bigg)\,.
 %\\
 %&=& m_c^{\overline{\mbox{\scriptsize MS}}}(\mu)\bigg(1 + C_F \frac{\alpha_s}{4\pi}
 %\bigg( 6\ln\left(\frac{\mu}{m_b}\right) - 3\ln\left(x_{+}\right) - 3\ln\left(x_{-}\right) + 4\bigg)\bigg)\,.
\end{eqnarray}
Note that LO-QCD diagrams not only contribute to this order, but 
also to NLO-QCD after replacing the bare bottom field and charm quark mass by their renormalized counterparts. 
%Note that, due to the 
%$\epsilon$ simple pole in the renormalization factors, the corresponding graph needs to be computed with $\mathcal{O}(\epsilon)$ accuracy.

\begin{figure}[!htb]  
	\centering
	\includegraphics[width=1.0\textwidth]{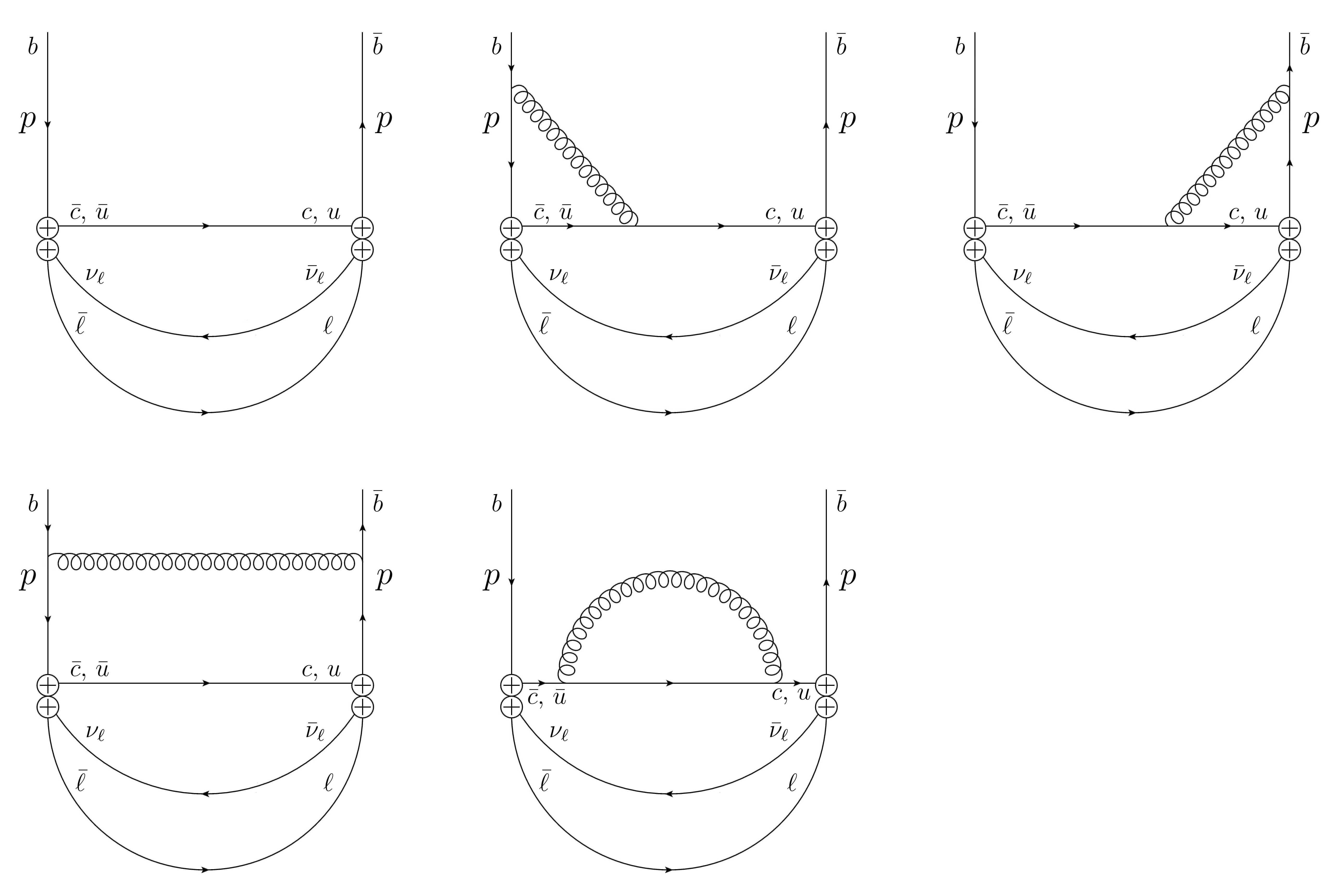}
        \caption{LO-QCD and NLO-QCD diagrams contributing to the total semi-leptonic decay width.}
        \label{WidthLOandNLO}
\end{figure}

\begin{figure}[!htb]  
	\centering
	\includegraphics[width=0.8\textwidth]{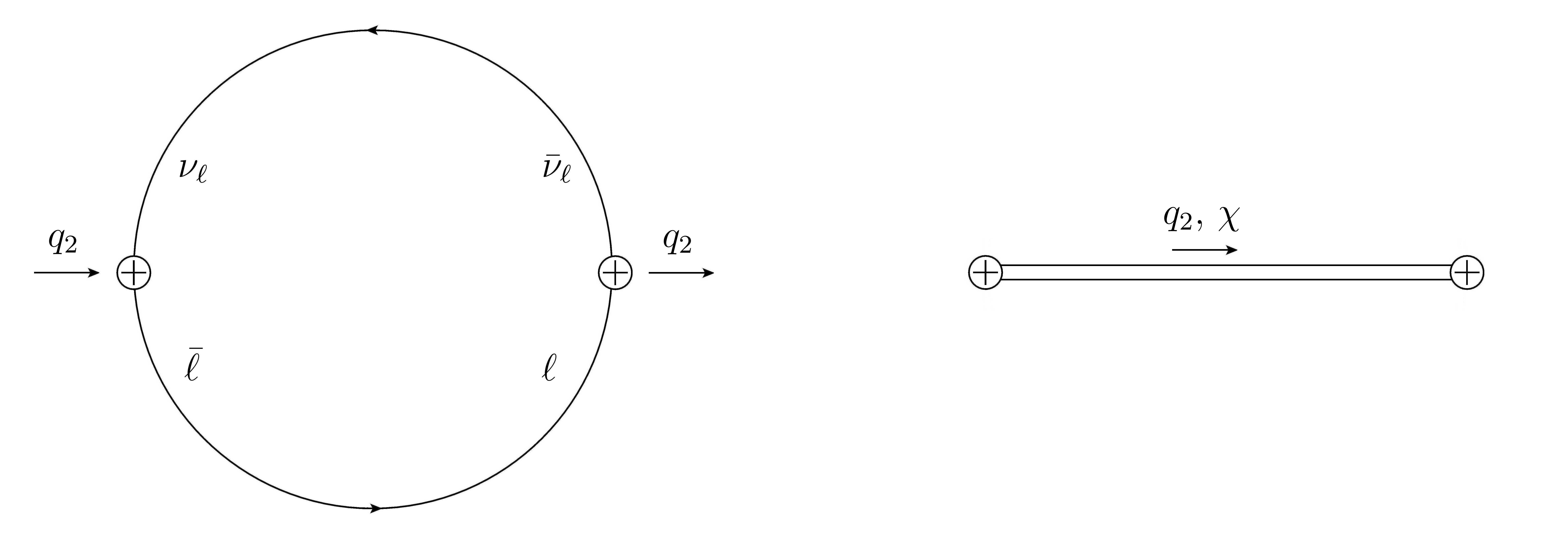}
        \caption{Graphs representing the leptonic loop appearing in the forward scattering matrix element of the $B$-hadron 
        in usual (left) and dispersion (right) representation.}
        \label{LeptonicLoop}
\end{figure}

The Feynman diagrams that contribute to the total semi-leptonic width at LO-QCD and NLO-QCD are the two-loop and three-loop diagrams shown 
in Fig.~[\ref{WidthLOandNLO}].
Following Eq.~(\ref{disp1LSm1m2Generic}), we use the dispersion representation of the one-loop sunset topology 
given in Eqs.~(\ref{disp1LSm1m2}) and (\ref{specden1LSm1m2}) to write the lepton-neutrino loop 
represented in Fig.[\ref{LeptonicLoop}] as an integral differential in the lepton pair invariant mass square $\chi^2$. 
That is, if both leptons are massless

\begin{equation}
\label{relrhosrho1LS}
\int\frac{d^D k}{(2\pi)^D}\frac{-\Tr(\Gamma^\sigma (\slashed k + \slashed q_2)\Gamma^\rho \slashed k)}{k^2(k+q_2)^2}  
  = i\int_{0}^{\infty}d(\chi^2)  \frac{\rho_s(\chi^2)}{\chi^2 - q_2^2-i\eta}(q_2^2 g^{\rho\sigma}-q_2^\rho q_2^\sigma)\,,
\end{equation}
where the spectral density $\rho_s$ reads

\begin{equation}
 \rho_s = \frac{D-2}{D-1}\rho_{{\scriptsize\mbox{1LS}}}(\chi^2,0,0) = 
 \frac{2}{3}\frac{1}{16\pi^2} + \mathcal{O}(\epsilon)\,.
\end{equation}
Since renormalization can be performed at the differential level, the integrand is finite, and it is enough to keep 
the $\mathcal{O}(\epsilon^0)$ term in $\rho_s$. 
A nice feature of the semi-leptonic width 
is that the purely leptonic part is not affected by QCD corrections, which keeps the spectral density very simple. 
Note that the ``effective massive propagator'' of mass $\chi$ is transverse due to the leptons are massless. 
After writing the leptonic loop in this form, we can compute the width differential 
in the dilepton pair invariant mass square by leaving the integral over
$\chi^2$ undone. 

\begin{figure}[!htb]  
	\centering
	\includegraphics[width=1.0\textwidth]{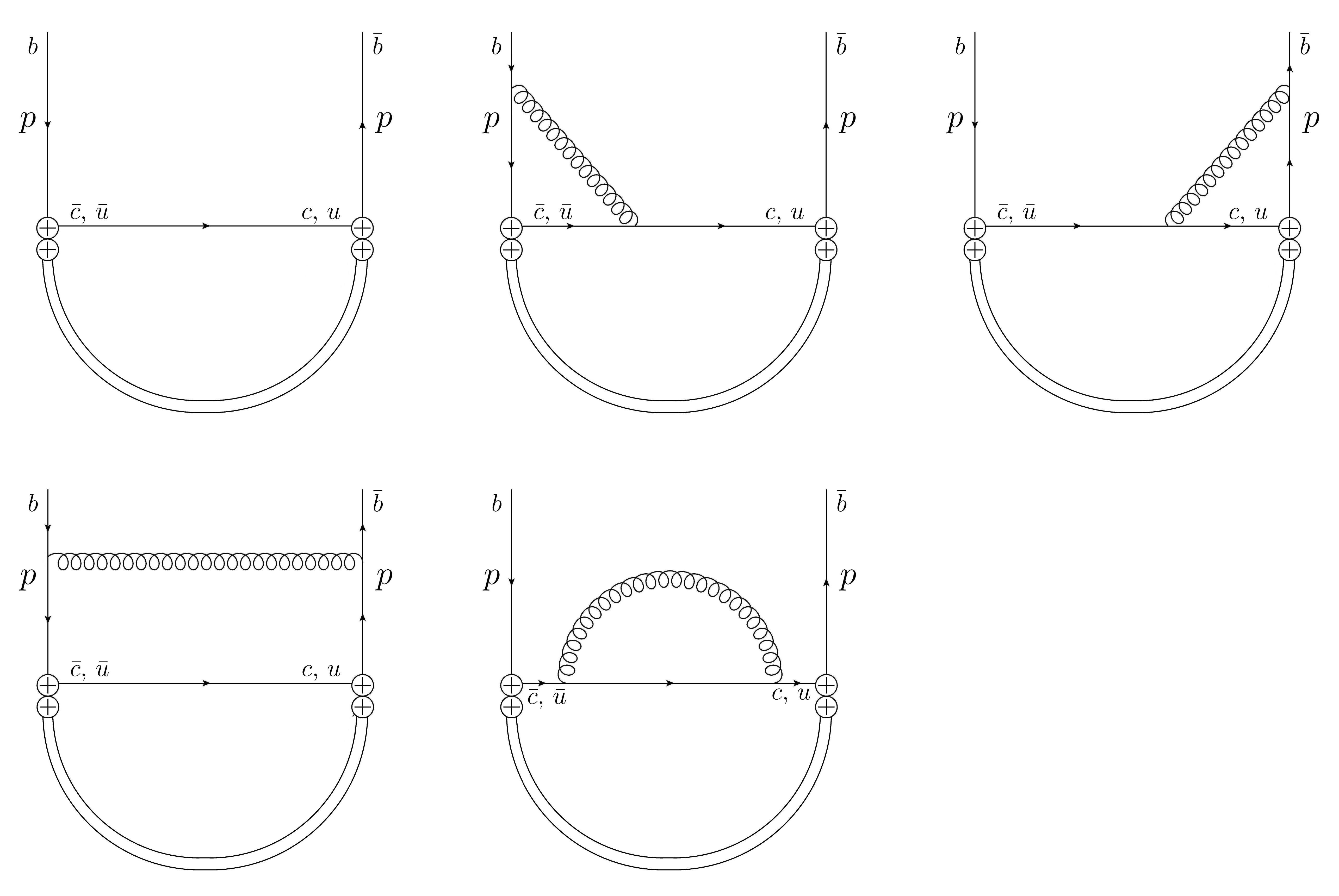}
        \caption{LO-QCD and NLO-QCD diagrams contributing to the differential semi-leptonic decay width.}
        \label{DifWidthLOandNLO}
\end{figure}

The Feynman diagrams contributing to the semi-leptonic differential width are shown in Fig.~[\ref{DifWidthLOandNLO}]. 
The corresponding amplitude is reduced to a combination of the master integrals computed in Sec.~\ref{masterint} by using LiteRed. 
In the $b\rightarrow c\ell\bar{\nu}_\ell$ channel, the amplitude at LO-QCD and NLO-QCD is written as a combination of the 
master integrals given in Secs. \ref{Sec:MILOQCDbclv} and \ref{Sec:MINLOQCDbclv}, respectively. 
In the $b\rightarrow u\ell\bar{\nu}_\ell$ channel, the amplitude at LO-QCD and NLO-QCD is written as a combination of the 
master integrals given in Secs. \ref{Sec:MILOQCDbulv} and \ref{Sec:MINLOQCDbulv}, respectively. 
Whereas the master integrals are universal objects, the particular combination that gives the semi-leptonic width 
depends on the structure of the interaction. 

The topology of the graphs is such that the NLO contribution can be only proportional to the color factor $C_F$.

The computation is done in a general covariant gauge with gauge fixing parameter $a$. The coefficients of the differential width 
are independent of the gauge fixing parameter, which is a strong check of the calculation.

Whereas the computation of $\alpha_s$ corrections to the semi-leptonic decay width stemming from the massive channel 
requires computing  three loop integrals with one scale ($\rho$), the computation of the 
differential width requires to evaluate two-loop integrals with two scales ($r$, $\rho$). That makes 
the analytical structure of the differential width more involved. 
Integrating over the additional scale $r$ one obtains the total width. 
Clearly, the additional integration makes the analytical structure of the total width simpler 
compared to its differential counterpart, since it is a function of one less parameter.

\section{Results}
\label{Sec:results}

In this section we summarize the analytical results for the HQE coefficients of the differential and total (integrated) semi-leptonic decay width 
stemming from both, the $b\rightarrow c\ell\bar{\nu}_\ell$ and the $b\rightarrow u\ell\bar{\nu}_\ell$ decay channels with massless leptons. 
Since we neglect the lepton masses, all $\eta_i$ vanish, so the 
differential rate is a function of $\rho$ and $r$, where $r$ is the invariant mass of the leptons. 
Integrating over $r$ yields the total rate, which then depends on $\rho$ only. 
  
For presentation, we split the LO and NLO contributions. We define the coefficients of the differential width to be

\begin{eqnarray}   
 \mathcal{C}_0 (\rho,r)  &=& \mathcal{C}_0^{\mbox{\scriptsize LO}} (\rho,r) + C_F \frac{\alpha_s}{\pi}\mathcal{C}_0^{\mbox{\scriptsize NLO}} (\rho,r) \,,
\end{eqnarray}
and the coefficients of the total (integrated) width to be

\begin{eqnarray}   
 C_0 (\rho) &=& C_0^{\mbox{\scriptsize LO}}(\rho) 
   + C_F \frac{\alpha_s}{\pi} C_0^{\mbox{\scriptsize NLO}}(\rho) \,.
\end{eqnarray}

%where $C_F=4/3$ is the quadratic Casimir operator of the fundamental representation of the $SU(3)$ group.

\subsection{Differential decay width}

For the $b\rightarrow c\ell\bar{\nu}_\ell$ channel the LO and NLO coefficients are most conveniently 
expressed in the variables $x_\pm$ and read 

\begin{eqnarray}
\mathcal{C}_0^{\mbox{\scriptsize LO}} &=& 
 48 \pi ^2 (x_{+}-x_{-}) \Big( x_{-} x_{+} (3 x_{-}+3 x_{+}-8) + x_{-}(3-2 x_{-}) + x_{+} (3-2 x_{+}) \Big)\,,
 \\
 \mathcal{C}_0^{\mbox{\scriptsize NLO}} &=& 
    -24\pi^2 \bigg\{
    \frac{1}{2}(x_{-}-x_{+})
   \nonumber
   \\
   && \quad
    \times \Big( x_{-} x_{+}(8 x_{-} x_{+} + 3 x_{-}  + 3 x_{+} - 28) + 3x_{-}(1 - 2x_{-}) + 3x_{+}(1 - 2x_{+}) + 8 \Big)
   \nonumber
   \\
   &&   
   + 2\Big(  x_{-} x_{+}( 3x_{-} + 3x_{+} - 8) + x_{-}(3 - 2x_{-}) + x_{+}(3 - 2x_{+}) \Big)
   \nonumber
   \\
   &&  \quad 
   \times\bigg[  
   (x_{-}+x_{+})\bigg( 2\Li_2\left(1-\frac{x_{-}}{x_{+}}\right) - 2\Li_2\left(1-\frac{x_{+}}{x_{-}}\right) + 2\Li_2(x_{-}) - 2\Li_2(x_{+})
   \nonumber
   \\
   && \quad
   + 2\ln (x_{-})\ln (x_{+}-x_{-}) - 2\ln(x_{+}) \ln(x_{+}-x_{-}) - \ln^2(x_{-}) + \ln^2(x_{+})
   \nonumber
   \\
   && \quad
   + \ln(1-x_{-})\ln (x_{-}) - \ln(1-x_{+}) \ln (x_{+})\bigg)
   -4 (x_{-}-x_{+})\ln (x_{+}-x_{-})\bigg]
   \nonumber
   \\
   &&
   + x_{-} \bigg(  x_{-}^2 x_{+}^2(4x_{+} - 14) + x_{-}x_{+}( -14x_{+}^2 + 34x_{+} + 26x_{-} - 40)
   \nonumber
   \\
   &&
                + x_{-}(15 - 12x_{-}) + x_{+}(5x_{+}^2 - 4x_{+} + 6) \bigg)\ln(x_{-})
   \nonumber
   \\
   &&
   - x_{+}\bigg( x_{-}^2 x_{+}^2(4x_{-} - 14) + x_{-}x_{+}( -14x_{-}^2 + 34x_{-} + 26x_{+} - 40)
   \nonumber
   \\
   &&
   + x_{+}(15 - 12x_{+}) + x_{-}(5 x_{-}^2 - 4x_{-} + 6) \bigg) \ln(x_{+}) 
   \nonumber
   \\
   &&
   - \bigg( 
    4x_{-}^3 x_{+}^3 
   - x_{-}^2 x_{+}^2(14x_{-} + 14x_{+} - 28)
   + x_{-} x_{+}( 14x_{+}^2 - 12x_{+} + 2x_{-}^2 + 12x_{-} - 28 )
   \nonumber
   \\
   &&
   + x_{-}( 4x_{-}^2 - 14x_{-} + 14) 
   + x_{+}(-4x_{+}^2 - 2x_{+} + 14) - 4 \bigg)\ln (1-x_{-})
   \nonumber
   \\
   &&
   +\bigg( 
   4x_{-}^3 x_{+}^3 
   - x_{-}^2 x_{+}^2 (14 x_{-} + 14x_{+} - 28)
   + x_{-}x_{+}(14 x_{-}^2 - 12x_{-} + 2x_{+}^2  + 12x_{+}  - 28)
   \nonumber
   \\
   &&
   + x_{+}( 4x_{+}^2 - 14x_{+} + 14 )
   + x_{-}(-4x_{-}^2 - 2x_{-} + 14) - 4\bigg) \ln(1-x_{+})
   \bigg\}\,.
\end{eqnarray}
Indeed, the terms involving $\Li_2(x)$ and $\ln^2(x)$ can be written more compactly using that
\begin{eqnarray}
   && 2\Li_2\left(1-\frac{x_{-}}{x_{+}}\right) - 2\Li_2\left(1-\frac{x_{+}}{x_{-}}\right) + 2\Li_2(x_{-}) - 2\Li_2(x_{+})
   + 2\ln (x_{-})\ln (x_{+}-x_{-})
   \nonumber
   \\
   &&
   - 2\ln(x_{+}) \ln(x_{+}-x_{-}) - \ln^2(x_{-}) + \ln^2(x_{+})
   + \ln(1-x_{-})\ln (x_{-}) - \ln(1-x_{+}) \ln (x_{+})
   \nonumber
   \\
    &=& \frac{\pi^2}{3} + 2\Li_2\left(1-\frac{x_{-}}{x_{+}}\right) -2\Li_2\bigg(\frac{x_{-}}{x_{+}}\bigg) 
   + \Li_2(x_{-}) - \Li_2(1-x_{-}) - \Li_2(x_{+}) + \Li_2(1-x_{+})
   \nonumber
   \\
   &=& 4\L\bigg(1-\frac{x_{-}}{x_{+}}\bigg) + 2\L(x_{-}) - 2\L(x_{+})\,,
   \label{identitydilog}
\end{eqnarray}
where $\L(x)$ is the Roger's dilogarithm~\cite{Kirillov:1994en}

\begin{equation}
 \L(x) = \Li_2(x) + \frac{1}{2}\ln(x)\ln(1-x) = \frac{1}{2}\bigg(\frac{\pi^2}{6} + \Li_2(x) - \Li_2(1-x) \bigg)
 ,\;\;\;\; 0<x<1\,.
\end{equation}
The second equality in Eq.~(\ref{identitydilog}) displays a explicit $\pi^2$ term which highlight the close 
relationship between factors of $\pi^2$ and dilogarithms. For example $\Li_2(1)=\pi^2/6$. As we will see throughout this section, 
we can always hide the explicit factors of $\pi^2$ by choosing a proper set of dilogarithms to express the results.

In terms of the Roger's dilogarithm 
the $\mathcal{C}_0^{\mbox{\scriptsize NLO}}$ coefficient reads

\begin{eqnarray}
 \mathcal{C}_0^{\mbox{\scriptsize NLO}} &=& 
    -24\pi^2 \bigg\{
    \frac{1}{2}(x_{-}-x_{+})
   \nonumber
   \\
   &&  \quad
    \times\Big( x_{-} x_{+}(8 x_{-} x_{+} + 3 x_{-}  + 3 x_{+} - 28) + 3x_{-}(1 - 2x_{-}) + 3x_{+}(1 - 2x_{+}) + 8 \Big)
   \nonumber
   \\
   &&   
   + 4\Big(  x_{-} x_{+}( 3x_{-} + 3x_{+} - 8) + x_{-}(3 - 2x_{-}) + x_{+}(3 - 2x_{+}) \Big)
   \nonumber
   \\
   && \quad
   \times\bigg[  
    (x_{-}+x_{+})\bigg( 2\L\bigg(1-\frac{x_{-}}{x_{+}}\bigg) + \L(x_{-}) - \L(x_{+}) \bigg)
  - 2(x_{-}-x_{+})\ln (x_{+}-x_{-}) \bigg]
   \nonumber
   \\
   &&
   + x_{-} \bigg(  x_{-}^2 x_{+}^2(4x_{+} - 14) + x_{-}x_{+}( -14x_{+}^2 + 34x_{+} + 26x_{-} - 40)
   \nonumber
   \\
   &&
                + x_{-}(15 - 12x_{-}) + x_{+}(5x_{+}^2 - 4x_{+} + 6) \bigg)\ln(x_{-})
   \nonumber
   \\
   &&
   - x_{+}\bigg( x_{-}^2 x_{+}^2(4x_{-} - 14) + x_{-}x_{+}( -14x_{-}^2 + 34x_{-} + 26x_{+} - 40)
   \nonumber
   \\
   &&
   + x_{+}(15 - 12x_{+}) + x_{-}(5 x_{-}^2 - 4x_{-} + 6) \bigg) \ln(x_{+}) 
   \nonumber
   \\
   &&
   - \bigg( 
    4x_{-}^3 x_{+}^3 
   - x_{-}^2 x_{+}^2(14x_{-} + 14x_{+} - 28)
   + x_{-} x_{+}( 14x_{+}^2 - 12x_{+} + 2x_{-}^2 + 12x_{-} - 28 )
   \nonumber
   \\
   &&
   + x_{-}( 4x_{-}^2 - 14x_{-} + 14) 
   + x_{+}(-4x_{+}^2 - 2x_{+} + 14) - 4 \bigg)\ln (1-x_{-})
   \nonumber
   \\
   &&
   +\bigg( 
   4x_{-}^3 x_{+}^3 
   - x_{-}^2 x_{+}^2 (14 x_{-} + 14x_{+} - 28)
   + x_{-}x_{+}(14 x_{-}^2 - 12x_{-} + 2x_{+}^2  + 12x_{+}  - 28)
   \nonumber
   \\
   &&
   + x_{+}( 4x_{+}^2 - 14x_{+} + 14 )
   + x_{-}(-4x_{-}^2 - 2x_{-} + 14) - 4\bigg) \ln(1-x_{+})
   \bigg\}\,.
\end{eqnarray}
Note that, in terms of $x_{\pm}$, the coefficient of the differential width can be expressed as a function of simple structures 
which are polynomials or polynomials multiplied by logarithms, square logarithms and dilogarithms. Alternatively, we can 
replace the dependence in square logarithms and dilogarithms in favour of Roger's dilogarithms.

It is also interesting to note the remarkable $x_{+}\leftrightarrow x_{-}$ asymmetry. More precisely, we observe that the quantity
$\mbox{Re}[\mathcal{C}_0(x_{+},x_{-}) + \mathcal{C}_0(x_{-},x_{+})]=0$. 

The coefficients for the $b\rightarrow u \ell\bar{\nu}_\ell$ channel can be obtained from the expressions for the 
$b\rightarrow c \ell\bar{\nu}_\ell$ channel by just taking $\rho=0$. In terms of the variables $x_{\pm}$, that corresponds to 
taking $x_{+} = 1-r$ and $x_{-}=0$. We check our results by doing both, taking the massless limit and computing them from scratch by using 
the master integrals in Sec. \ref{Sec:MILOQCDbulv} and \ref{Sec:MINLOQCDbulv}. They read

\begin{eqnarray}
  \mathcal{C}_0^{\mbox{\scriptsize LO}}  &=& 48 \pi ^2 (1-r)^2 (1+2r) \,,
  \\
  \mathcal{C}_0^{\mbox{\scriptsize NLO}} 
  %&=& 4\pi^2 \bigg[  3(1-r)(6r^2-9r-5) + 12r (1+r)(1-2r) \ln(r) 
  %\nonumber
   %\\
   %&&
   %+  6(1-r)^2(5 + 4r) \ln(1-r)
  %\nonumber
  %\\
  %&&
  %+ 24(1-r)^2(1+2r) \bigg( \frac{\pi^2}{6} + \frac{1}{2}\ln(1-r)\ln(r) + \Li_2(r) \bigg)
  %\bigg]
  %\\
  &=& -4\pi^2 \bigg[ 
    3(1-r)(6r^2-9r-5) + 12r (1+r)(1-2r) \ln(r) 
   \nonumber
   \\
   &&
   +  6(1-r)^2(5 + 4r) \ln(1-r)
  \nonumber
  \\
  &&
  + 12(1-r)^2(1+2r) \bigg( \frac{\pi^2}{2}  + \Li_2(r) - \Li_2(1-r) \bigg)
   \bigg]
   \label{C0widthmassless1}
   \\
  &=& -4\pi^2 \bigg[ 
   3(1-r)(6r^2-9r-5) + 12r (1+r)(1-2r) \ln(r) 
   \nonumber
   \\
   &&
   +  6(1-r)^2(5 + 4r) \ln(1-r)
  \nonumber
  \\
  &&
  + 24(1-r)^2(1+2r) \bigg( 2\L(r) + \L(1-r) \bigg)
   \bigg]\,.
   \label{C0widthmassless2}
\end{eqnarray}
Note that the combination $\Li_2(r) - \Li_2(1-r)$ do not produce any factor of $\pi^2$ after integration, so to the total width. 
That means that the $\pi^2$ term in the coefficient of the total width comes only from the integration of the term with 
explicit $\pi^2$ in Eq.~(\ref{C0widthmassless1}). 
In Eq.~(\ref{C0widthmassless2}), the $\pi^2$ terms are completely hidden inside Roger's dilogarithms.

\subsection{Total (integrated) decay width}

Once the coefficient of the differential semi-leptonic decay width is obtained, we can compute the coefficient of the total semi-leptonic decay width 
as a check. It is obtained by solving the following integral
\begin{equation}
 C_0(\rho) = \frac{1}{24\pi^2} \int_{0}^{(1-\sqrt{\rho})^2}d r\,\mathcal{C}_0(r,\rho)\,.
\end{equation}
Note that the expression above requires an integration over $r$. 
However, our expression for $\mathcal{C}_0(r,\rho)$ is written in terms of $x_{\pm}(r,\rho)$, which are the most natural variables 
for describing the system. As we previously discussed, these variables keep the expressions for the coefficients rather compact and in terms of simple structures which are 
polynomials in $x_{\pm}$, or polynomials multiplied by logarithms, square logarithms and dilogarithms. 
On the contrary, if we rewrite the coefficient of the differential width in terms of $r$ and $\rho$, 
the structure turns out to be much more complicated and integration becomes non-viable. 
Therefore, it is much better to integrate over $x_{+}$ or $x_{-}$ instead of $r$. First we use 
$x_{+} = \rho/x_{-}$ in order to write $\mathcal{C}_0$ in terms of a single variable $x_{-}$ ($\rho$ is constant). Then, 
$dr =(-1+\rho/x_{-}^2)dx_{-}$ and the integral becomes 

\begin{equation}
\label{intcoefwidth}
 C_0(\rho) = \frac{1}{24\pi^2} \int_{\sqrt{\rho}}^{\rho}dx_{-} 
 \bigg(1 - \frac{\rho}{x_{-}^2}\bigg) \mathcal{C}_0\left(x_{+}=\frac{\rho}{x_{-}},x_{-}\right)\,.
\end{equation}
After integration we find the coefficients for the $b\rightarrow c\ell\bar{\nu}_\ell$ channel

\begin{eqnarray}
 C_0^{\mbox{\scriptsize LO}} &=& 1 -8\rho + 8\rho^3 -\rho^4 - 12\rho^2 \ln (\rho) \,,
 \\
 C_0^{\mbox{\scriptsize NLO}} &=& 
 \frac{25}{8} - \frac{\pi ^2}{2} -\frac{239 \rho}{6} + 16 \pi ^2 \rho ^{3/2} -8 \pi ^2 \rho ^2 + 16 \pi ^2 \rho ^{5/2}
 +\frac{239 \rho^3}{6} -\frac{25 \rho ^4}{8} -\frac{\pi ^2\rho ^4}{2}
 \nonumber
 \\
 &&
 -\frac{17}{6} \ln (1-\rho) +\frac{32}{3} \rho  \ln (1-\rho) -\frac{32}{3} \rho ^3 \ln (1-\rho)  + \frac{17}{6} \rho ^4 \ln (1-\rho)
  \nonumber
 \\
 &&
 -10 \rho  \ln (\rho) - 45 \rho^2 \ln (\rho) + \frac{2}{3} \rho ^3\ln (\rho) - \frac{17}{6} \rho ^4 \ln (\rho)
 \nonumber
 \\
 &&
 - 32 \rho^{3/2} \ln \left(1-\sqrt{\rho }\right) \ln (\rho)   
 - 32\rho ^{5/2} \ln \left(1-\sqrt{\rho }\right) \ln (\rho)
 \nonumber
 \\
 &&
 + 32 \rho ^{3/2} \ln \left(1+\sqrt{\rho }\right) \ln (\rho) 
 + 32 \rho ^{5/2} \ln \left(1+\sqrt{\rho }\right) \ln (\rho)
 \nonumber
 \\
 &&
 + 2\ln (1-\rho)\ln(\rho) + 60\rho^2 \ln (1-\rho) \ln (\rho) + 2 \rho^4 \ln (1-\rho)\ln(\rho)
 \nonumber
 \\
 &&
 - 18 \rho^2 \ln ^2(\rho) -\frac{1}{2}\rho ^4 \ln^2(\rho)
 \nonumber
 \\
 &&
 + (6 + 64\rho^{3/2} + 96\rho^2 + 64\rho^{5/2} + 6\rho^4) \Li_2(-\sqrt{\rho })
 \nonumber
 \\
 &&
 + (6 - 64\rho^{3/2} + 96\rho^2 - 64\rho^{5/2} + 6\rho^4) \Li_2(\sqrt{\rho })\,,  
 \label{c0width1}
\end{eqnarray}
which are in agreement with previously known results given in Ref.~\cite{Nir:1989rm,Mannel:2014xza,Mannel:2015jka}.
Indeed, the expression above can be written more compactly by choosing a different set of dilogarithms to express the result 

\begin{eqnarray}
 C_0^{\mbox{\scriptsize NLO}} &=& 
128 \rho ^{3/2} (\rho +1) \Li_2\left(1-\sqrt{\rho }\right)-\left(32 \rho ^{5/2}+32 \rho ^{3/2}+3 \rho ^4+48 \rho ^2+3\right)
   \Li_2(1-\rho )
   \nonumber
   \\
   &&
   -\frac{1}{2} \rho ^2 \left(\rho ^2+36\right) \ln ^2(\rho )+\frac{1}{24} \left(-75 \rho ^4+956 \rho ^3-956
   \rho +75\right)
   \nonumber
   \\
   &&
   +\frac{1}{6} \left(17 \rho ^4-64 \rho ^3+64 \rho -17\right) \ln (1-\rho )-\left(\rho ^4-12 \rho ^2+1\right)
   \ln (1-\rho ) \ln (\rho )
   \nonumber
   \\
   &&
   -\frac{1}{6} \rho  \left(17 \rho ^3-4 \rho ^2+270 \rho +60\right) \ln (\rho )\,,
   \label{c0width2}
\end{eqnarray}
or even in terms of Roger's dilogarithm

\begin{eqnarray}
 C_0^{\mbox{\scriptsize NLO}} &=& 
 128 \rho ^{3/2} (\rho +1) \L\left(1-\sqrt{\rho }\right)-\left(32 \rho ^{5/2}+32 \rho ^{3/2}+3 \rho ^4+48 \rho ^2+3\right) \L(1-\rho
   )
      \nonumber
   \\
   &&
   -32 \rho ^{3/2} (\rho +1) \ln \left(1-\sqrt{\rho }\right) \ln (\rho )-\frac{1}{2} \rho ^2 \left(\rho ^2+36\right) \ln
   ^2(\rho )
      \nonumber
   \\
   &&
   +\frac{1}{24} \left(-75 \rho ^4+956 \rho ^3-956 \rho +75\right)+\frac{1}{6} \left(17 \rho ^4-64 \rho ^3+64 \rho
   -17\right) \ln (1-\rho )
      \nonumber
   \\
   &&
   -\frac{1}{6} \rho  \left(17 \rho ^3-4 \rho ^2+270 \rho +60\right) \ln (\rho )
   \nonumber
   \\
   &&
   +\frac{1}{2} \left(32
   \rho ^{5/2}+32 \rho ^{3/2}+\rho ^4+72 \rho ^2+1\right) \ln (1-\rho ) \ln (\rho )\,.
   \label{c0width3}
\end{eqnarray}
Note that in both cases explicit $\pi^2$ terms are hidden inside dilogarithms. 

For the $b\rightarrow u\ell\bar{\nu}_\ell$ channel, the integration of Eq. (\ref{intcoefwidth}) over $r$ can be easily performed since $\rho=0$. 
Alternatively, we can take the massless limit in Eqs. (\ref{c0width1})-(\ref{c0width3}). In both cases we obtain 
agreement with the known result

\begin{equation}
 C_0^{\mbox{\scriptsize LO}} = 1\,,\quad\quad 
 C_0^{\mbox{\scriptsize NLO}} = \frac{1}{8} \left(25-4 \pi ^2\right)\,.
\end{equation}

\section{Conclusions}

We have presented an alternative method for the computation of the decay width differential in the $n$ $p$-particle invariant mass square 
at NLO in the strong coupling $\alpha_s$, assuming a general transition $Q(M)\rightarrow q(m)p_1(m_1)\ldots p_n(m_n)$. 
The approach is based on the use of the optical theorem, the HQE and dispersion representations. 
The main achievement of this paper is the computation of the associated master integrals analytically, which are 
universal to this kind of processes. We have observed that the key point for the analytical computation of the master integrals in the massive 
case ($m\neq 0$) is the choice of the most natural variables for the description of the system. We also compute the master 
integrals in the massless ($m=0$) case as a check.

Note that the coefficient of the differential width can be always computed analytically. 
Being able to get a completely analytical result or not will depend on the complexity of the spectral density $\rho_s$.
If $\rho_s$ can be computed analytically, then the differential width is known analytically and the final integration over 
$r$ to obtain the total width or moments (with or without cuts) can be always performed, at least numerically, which is enough for comparison to 
experiment.

As an application, we have computed the leading HQE coefficient of the semi-leptonic decay width differential in the dilepton pair invariant mass square 
at NLO in $\alpha_s$. The decay is mediated by the $b\rightarrow q\ell\bar{\nu}_\ell$ transition, where the leptons are considered to be massless 
and $q$ can be either massive or massless. By a clever choice of variables, the differential width 
for the massive case can be expressed in terms of simple structures which 
are polynomials and polynomials multiplied by logarithms and Roger's dilogarithms (or alternatively polynomials multiplied by 
square logarithms and dilogarithms). 
%The result we present is new. 
The differential width for the massless case is computed as a check of our methods. We compute it explicitly and 
compare with the massive case in the limit $m \to 0$. 
As an additional check, we integrate the differential width to get the total semi-leptonic width and compare with the known result.

The method we use together with the master integrals we have computed in this paper can be straightforwardly 
extended for the analytical computation of the power corrections of the HQE at NLO-QCD.
We plan to address this computation in future publications. 

Let us stress that the master integrals we have computed may have many physical applications. They can be used for the 
computation of the differential and total semi-leptonic and non-leptonic decay widths with any number of different masses, 
or to topologically analogous decays emerging from new physics interactions.

%The computation of moments of the distribution in different kinematic invariants with or without cuts is also possible. 
%For semi-leptonic and non-leptonic decays, numerical integration to obtain the total width or moments is always possible since the 
%coefficient of the width is always known analytically and $\rho_s$ can be obtained analytically by using the dispersion representation 
%of the one-loop sunset topology.

\subsection*{Acknowledgments}
This research was supported by the Deutsche Forschungsgemeinschaft 
(DFG, German Research Foundation) under grant  396021762 - TRR 257 
``Particle Physics Phenomenology after the Higgs Discovery''. 

\begin{appendices}
\section*{Appendix}
\label{sec:appendix}

\subsection*{Dispersion representation of the one-loop two-propagator diagram}

Two-point functions obey dispersion relations which follow from the analyticity 
properties of these functions~\cite{Kallen:1952zz,Lehmann:1954xi}. In particular, the 
dispersion representation of the one-loop diagram with two propagators  
plays a key role in this work. It is not only extremely useful for the computation 
of master integrals, but most importantly it allows to write the expression for the semi-leptonic 
width in such a way that the 
decay width differential in the leptonic pair invariant mass can be identified. 
For this reason, we present it here explicitly due to its relevance to the paper. 

In the case of two massive lines with different mass the dispersion representation of the one-loop sunset reads

\begin{equation}
 \label{disp1LSm1m2}
 \frac{1}{i}\int\frac{d^D q}{(2\pi)^D}\frac{1}{q^2-m_1^2}\frac{1}{(p-q)^2 - m_2^2}
 = \int_{(m_1+m_2)^2}^{\infty}  \frac{\rho_{{\scriptsize\mbox{1LS}}}(\xi,m_1,m_2)}{\xi - p^2-i\eta}d\xi\,,
\end{equation}
where

\begin{equation}
\label{specden1LSm1m2}
\rho_{{\scriptsize\mbox{1LS}}}(\xi,m_1,m_2) = \frac{1}{(4\pi)^{\lambda+1}}\frac{\Gamma(\lambda)}{\Gamma(2\lambda)}
\xi^{\lambda-1}\bigg[ \bigg(1- \frac{(m_1+m_2)^2}{\xi}\bigg)
\bigg(1-\frac{(m_1-m_2)^2}{\xi}\bigg)\bigg]^{\lambda-1/2}\,,
\end{equation}
with $D=2\lambda+2 = 4-2\epsilon$, is the spectral density.

\end{appendices}

\clearpage

\newpage

\end{document}